\documentclass[iop,twocolumn,twocolappendix]{emulateapj}
\usepackage{natbib}
\usepackage{amsmath}
\usepackage{epstopdf}
\usepackage{gensymb}

\usepackage{color}
\usepackage{soul}

\newcommand{\red}[1]{#1}

\begin{document}
\title{Axisymmetric Simulations of Hot Jupiter-Stellar Wind Hydrodynamic Interaction }

\author{Duncan Christie, Phil Arras, and Zhi-Yun Li}
\affil{Department of Astronomy, University of Virginia}
\begin{abstract}

Gas giant exoplanets orbiting at close distances to the parent star are subjected to large radiation and stellar wind fluxes. In this paper, hydrodynamic simulations of the planetary upper atmosphere and its interaction with the stellar wind are carried out to understand the possible flow regimes and how they affect the Lyman $\alpha$ transmission spectrum. Following Tremblin and Chiang, charge exchange reactions are included to explore the role of energetic atoms as compared to thermal particles. In order to understand the role of the tail as compared to the leading edge of the planetary gas, the simulations were carried out under axisymmetry, and photoionization and stellar wind electron impact ionization reactions were included to limit the extent of the neutrals away from the planet. By varying the planetary gas temperature, two regimes are found. At high temperature, a supersonic planetary wind is found, which is turned around by the stellar wind and forms a tail behind the planet. At lower temperatures, the planetary wind is shut off when the stellar wind penetrates inside where the sonic point would have been. In this regime mass is lost by viscous interaction at the boundary between planetary and stellar wind gases. Absorption by cold hydrogen atoms is large near the planetary surface, and decreases away from the planet as expected. The hot hydrogen absorption is in an annulus and typically dominated by the tail, at large impact parameter, rather than by the thin leading edge of the mixing layer near the substellar point.

\end{abstract}

\section{Introduction}
\label{sec:Introduction}

Hot Jupiters are planets with masses comparable to that of Jupiter, and orbital radii within $0.1\,{\rm AU}$ of their host stars. Due to the high levels of irradiation and stellar wind flux, the possibility exists for these planets to lose mass and form a stream of gas trailing the planet, analogous to sun-grazing comets \citep{1998ASPC..134..241S}. This paper studies hydrodynamic escape from the planet, interaction with the stellar wind, and the observability of the hydrogen atoms which originated from the planet, but some of which may undergo charge exchange with hot protons in the stellar wind.

The detection of extended hydrogen atmospheres around close-in exoplanets has relied primarily on Lyman-$\alpha$ absorption during transit, subject to the constraint that absorption by the interstellar medium (ISM) and geocoronal emission prevent any observation within $\sim 50\, {\rm km\, s^{-1}}$ of line center.    

The first detections of Lyman-$\alpha$ absorption by HD 209458b found  a transit depth of $15\pm 4\%$ \citep{2003Natur.422..143V}, primarily in the blue wing, supporting the idea of an extended, escaping hydrogen atmosphere surrounding the planet.  Subsequent observations of HD 209458b by \citet{2008ApJ...688.1352B} found significantly less absorption, only $8.9\pm 2\%$, with absorption in both the red and blue wings. A more natural explanation in the absence of blue-shifted absorption is a large column of static gas absorbing out on both the red and blue wings of the line. Lyman-$\alpha$ absorption has similarly been observed during the transit of HD 189733b \citep{2010A&A...514A..72L}, and as with HD 209458b, subsequent observations found differing results \citep{2013A&A...551A..63B}.   Fluctuations in stellar EUV and X-ray activity, drivers of the mass loss, can increase or decrease the size of the hydrogen cloud surrounding the planet, thus changing the Lyman-$\alpha$ transit depth.   Additionally, due to stellar variability in the UV across the disk of the star, transits observed in these wavelength ranges also exhibit much larger variation in observed transit depths compared to those observed in the optical \citep{2015ApJ...802...41L}. Recently, a 50\% transit depth mainly on the blue wing of the line has been observed for the Neptune-sized planet GJ 436b \citep{2015Natur.522..459E}, which is a clear sign of absorption by fast atoms moving away from the planet and toward the observer.

Theoretical models attempting to understand Lyman-$\alpha$ absorption are primarily divided into two groups: those that treat the gas primarily as a fluid and focus on the deeper regions of the atmosphere and those that focus on more rarefied regions where the gas originating from the planet interacts with the stellar wind.

One dimensional models of the atmospheres of hot Jupiters have found that EUV heating is capable of launching a supersonic wind which results in significant neutral hydrogen density out to several planetary radii \citep{2004Icar..170..167Y,2009ApJ...693...23M,2013Icar..226.1678K,2013Icar..226.1695K}.  In the deepest reaches of this neutral layer, the column of hydrogen is sufficient to absorb Lyman-$\alpha$ photons away from line center, potentially explaining the observed transit depth.    

In the other class of models, hydrogen atoms are launched from the exobase above the planetary photosphere \footnote{ The abundance of hydrogen atoms in the $\sim 10^6\ {\rm K}$ stellar wind is negligible, hence the source of atoms must be the planet.} and move under the influence of gravity, radiation pressure, and collisions with stellar wind protons. Radiation pressure arising from stellar Lyman $\alpha$, if optically thin, is capable of accelerating hydrogen to speeds $\sim 200\, {\rm km\, s^{-1}}$ away from the star \citep{2003Natur.422..143V}.   Charge-exchange interactions between neutral planetary hydrogen and stellar wind protons result in energetic neutral atoms (ENAs) with velocities characteristic of the stellar wind \citep{2008Natur.451..970H}.  Due to its broad linewidth, this population of fast-moving hydrogen has a higher scattering cross section at $\ga 100\ {\rm km\ s^{-1}}$ from line center. In these models neutral hydrogen is launched ballistically away from the planet with a density and speed taken from hydrodynamic simulations.

The point of departure for the present study is \citet{2013MNRAS.428.2565T}, who studied the hydrodynamic interaction of the planetary wind with the stellar wind. They considered separate populations of hot ($T \sim 10^6\,{\rm K}$) and cold ($T\sim 10^4\,{\rm K}$) hydrogen atoms and protons, which are advected with the flow, and allowed to interact by charge exchange reactions.   The planetary wind was launched with density and temperature taken from the hydrodynamic simulations of \citet{2009ApJ...693...23M}, and it expands outward until it meets the stellar wind. It is within the unstable mixing layer between the two gases that the hot neutral population is found.  

\citet{2013MNRAS.428.2565T} found that the hot hydrogen was capable of explaining the observed Lyman-$\alpha$ absorption at $\ga 100\, {\rm km\, s^{-1}}$, which is only $\sim 1-2$ Doppler widths from line center for the hot population of hydrogen, and hence requires much smaller columns since the cross section is much larger than that of $T \sim 10^4\ {\rm K}$ gas. Due to the setup for their simulations, the focus was on absorption by gas on the upstream side of the planet, where the stellar and planetary winds first interact.

In this paper, we follow \citet{2013MNRAS.428.2565T} to consider hydrogen gas driven from the planetary surface, the hydrodynamic interaction with the stellar wind, and charge exchange reactions in the mixing layer. We explore the resultant spatial distributions of the stellar wind and planetary gases and show there are two regimes:  colliding winds, as discussed by \citet{2013MNRAS.428.2565T}, and a ``viscous" interaction regime, in which the supersonic planetary wind is suppressed by the interaction with the stellar wind. The simulations are carried out in spherical coordinates, assuming axisymmetry, allowing the study of a tail formed by interaction with the high speed stellar wind. Photo- and collisional ionization are included in our chemical model, which act to limit the neutral population far from the planet.

\red{A number of important physical effects have been neglected in this paper, including stellar tides, stellar and planetary magnetic fields, sub- versus super-Alfv\'enic wind speeds, and three dimensional effects from stellar irradiation.  Some of these effects are treated in, e.g.,  \citet{2009ApJ...693...23M,2011MNRAS.411L..46V,2011ApJ...733...67C,2011ApJ...738..166C,2013ApJ...764...19B,2015A&A...578A...6M}.}

The paper is organized as follows: \S \ref{sec:outline} contains the physical assumptions motivating the numerical simulations to be 
performed. \S \ref{sec:numericalmethod} discusses the numerical method for the simulations. \S \ref{sec:hotcolddist} discusses how the hot and cold populations are distributed and their ionization state. \S \ref{sec:mlr} discusses the influence of the stellar wind in regulating the mass loss rate. \S \ref{sec:mfp} discusses the mean free paths and the validity of the fluid approximation. \S \ref{sec:lymanalpha} presents calculations of Lyman-$\alpha$ absorption as a function of geometry and the binding energy of the gas. 

\S \ref{sec:discussion} is the final summary and discussion.

\section{ Outline of the Problem }
\label{sec:outline}

We consider a planet of mass $M_{\rm p}$ and broadband optical
radius $R_{\rm p}$ in a circular orbit of radius $D$ around the
parent star. Energetic radiation from the star will heat the planet's
upper atmosphere to a temperature $T_{\rm p}$ which is typically
much higher than the temperature of the optical or infrared
photospheres. The main parameter for understanding the hydrodynamic
outflow from the upper atmosphere of the planet is the ratio of the
particle's binding energy to the thermal energy \footnote{ The parameter $\lambda$
is also used later for the wavelength, but this should cause no confusion.},
\begin{eqnarray}
\lambda & =& \frac{GM_{\rm p}\mu m_{\rm u}/R_{\rm p} }{k_{\rm B} T_{\rm p} }
= \frac{ GM_{\rm p} }{ R_{\rm p} c_{\rm p}^2 }.
\label{Eqn:EscapeParam}
\end{eqnarray}
Here $G$ is the gravitational constant, $k_{\rm B}$ is Boltzmann's
constant, $m_{\rm u}$ is the atomic mass unit, $\mu m_{\rm u}$ is
the mean molecular weight of the gas, and $c_{\rm p} = ( k_{\rm B}
T_{\rm p} / \mu m_{\rm u} )^{1/2}$ is the isothermal sound speed.  \red{Up to a factor of two, the escape parameter is also the ratio of the escape speed for the planet squared to the sound speed squared.}

Ignoring for the moment the role of the stellar wind, the planet
will have a vigorous outflow for small $\lambda$ , and
exponentially decreasing mass loss as $\lambda$ increases \footnote{
At sufficiently large $\lambda$, the gas will become collisionless
below the sonic point, and the hydrodynamic treatment is no longer
valid \citep{2011ApJ...729L..24V}.}  (see eq.\ref{eqn:masslosswind}).
Larger $T_{\rm p}$ (smaller $\lambda$) also increases the scale
height and gas density with altitude, leading to increased absorption
of stellar flux by the atmosphere during transits. In the isothermal
limit, the outflow accelerates to the sound speed at the sonic point
radius
\begin{eqnarray}
r_{\rm sonic} & = & \frac{G M_{\rm p}}{ 2 c_{\rm p}^2 } = \frac{\lambda}{2} R_{\rm p},
\label{eq:rsonic}
\end{eqnarray}
with supersonic flow and density $\rho_{\rm p} \propto r^{-2}$
outside this point and a nearly hydrostatic profile inside.

Including the stellar wind, the planetary outflow is confined by
the wind's ram pressure. For a strong planetary outflow and weak
stellar wind, the boundary between the two fluids (contact
discontinuity) is far from the planet, while for a weak planetary
outflow and strong stellar wind it approaches the planet. In \S
\ref{sec:mlr} it is shown that the planetary wind is quenched
when the stellar wind can penetrate all the way down to the sonic
point of the planetary wind.

Hence there are two regimes for the interaction of the planetary
and stellar wind gases. For small $\lambda$ and a weak stellar wind,
the boundary between the two fluids is outside the sonic point, and
a transonic planetary wind meets the supersonic stellar wind. This
will be referred to as the ``colliding winds" regime. In the opposite
limit of large $\lambda$ and strong stellar wind, the stellar wind
penetrates deeper than where the planet's sonic point would have
been, and the net result is the stellar wind meeting a nearly
hydrostatic planetary atmosphere. This will be referred to as the
``viscous" regime, in which mass loss is due to friction at the
unstable interface of the two fluids, where the stellar wind entrains 
the planetary gas through turbulent mixing.

\red{It should be noted that while we present variations in $\lambda$ as variations
in the temperature of the planetary outflow, variations in $\lambda$ can also be viewed
as changes in the mass and radius of the planet.  In Section 4 it is shown that for a transonic planetary outflow to occur, the planetary ram pressure must hold the stellar wind outside the sonic point of the planetary outflow. Hence the important parameter to determine if a transonic planetary outflow can occur is the ratio of planetary ram pressure to stellar wind ram pressure, evaluated at the sonic point of the planetary outflow. To illustrate the two flow regimes, here the binding parameter $\lambda$ is varied. Equivalently, $\lambda$ could have been held constant and the stellar wind density and velocity could have been varied. See Section 5 for an analytic estimate of the critical value of $\lambda$ required for a transonic outflow, given the planetary mass, radius and base density, and the stellar wind density and velocity.
}

The observed transit depth, or fractional drop in flux during
transit, is roughly the fractional area of the star covered by
optically thick lines of sight through the neutral hydrogen gas surrounding the 
planet.
Two populations of hydrogen (``cold" and ``hot") contribute to the flux drop during transit\footnote{\red{When referring to {\em hydrogen} throughout the remainder of the text, we intend it to mean hydrogen in the neutral state.  Ionized hydrogen will be referred to as {\em protons}.}}.

Both populations of atoms originate in the planet, with cold hydrogen having temperature
characteristic of the planetary gas, while the hot population is due to a 
charge exchange reactions between
cold atoms from the planet and solar wind protons.

The temperature of the cold hydrogen atoms escaping from the planet
is regulated to be $T_{\rm p} \sim 10^4\ {\rm K}$ by a balance of
photoelectric heating, collisionally excited Lyman-$\alpha$ emission
and adiabatic cooling \citep{2009ApJ...693...23M}.  The thermal
Doppler width of these atoms, expressed in velocity units, is then
$ c_p \sim 10\ {\rm km\ s^{-1}}$. To affect the part of the Lyman
$\alpha$ line profile at velocities $ \geq 100\ {\rm km\ s^{-1}}$
from line center, where absorption by the interstellar medium (ISM)
is not complete, there must be a sufficiently large column of cold
hydrogen to make the gas optically thick at $\sim 10$ Doppler widths
from line center. However, since the cross section at 10 Doppler
widths is $\sim 10^{-6}$ times smaller than the line center cross
section, an optical depth of unity requires quite large columns of
cold hydrogen atoms covering a significant fraction of the disk of
the star. The required hydrogen columns would be much reduced 
if the planetary gas could be entrained and accelerated to the 
speed of the stellar wind, which is comparable to its sound speed 
$c_\star$, or if the gas had a larger thermal speed.

A possible source of hydrogen atoms with larger thermal speeds is
through charge exchange of a thermal hydrogen atom from the planet
at $T_{\rm p}\sim 10^4\ {\rm K}$ with a stellar wind proton at
$T_\star \sim 10^6\ {\rm K}$, creating a hydrogen atom with temperature
$\sim T_\star$.

Since the hot hydrogen population 
has a thermal Doppler width of $c_\star \sim\ 100\ {\rm km\ s^{-1}}$,
it can produce the observed absorption near line center, with a far
smaller column than that required of the cold hydrogen population.

The absorption by the cold and hot populations of hydrogen atoms
are not independent; the number of hot atoms depends on the density
of cold planetary gas at the position of the contact discontinuity
with the stellar wind.  For instance, it is possible that optical depth $\tau \sim 1$ for
the hot population would imply $\tau \gg 1$ for the cold population, and the model
transit depth would overestimate the observed transit depth when both contributions
are included. Hence a self-consistent model for the source of atoms,
as discussed in this paper, is required to understand the combined effect of both
populations.

The spatial distributions of the cold and hot populations are also
different. The cold population is strongly centered on the planet,
as the optical depths through the deep atmosphere are enormous. For 
example, the line center Lyman $\alpha$ optical depth for a hydrogen
layer with base pressure $1\ \mu {\rm bar}$ is $\tau \sim 10^8$. By 
contrast, the
hot population traces out the mixing layer between the planetary
and stellar wind gases. As shown in \S \ref{sec:hotcolddist},
the spatial distribution of each population depends on the regime of
the hydrodynamic outflow. In the colliding winds regime, the hot
population forms a thin sheath draping around the outside of the
planetary gas, well away from the planet, with a size comparable
or larger than the stellar disk.  In the viscous regime, by contrast,
the hot population mixes into a thin tail comparable to the size
of the planet, and smaller than the stellar disk.  In the
absence of photo- and collisional ionization by the stellar
wind, the ionization state would be frozen into the outflowing gas
from the planet, and for a fixed cross sectional area of the tail,
large optical depths could be obtained even far from the planet.
For the stellar wind parameters used here, collisional ionization 
by stellar wind electrons dominates photoionization, and hence 
the planetary gas becomes highly ionized once it mixes sufficiently
with the stellar wind. The photoionization rate, while an order of magnitude lower,
acts even when the planetary and stellar gases have not become well mixed.

In regards to charge exchange reactions, the hydrodynamic model
used in this paper assumes that the mean free path is smaller than
other lengthscales in the problem. This assumption will be checked
a posteriori (see Figure \ref{Fig:CEMFP} and the associated discussion in \S \ref{sec:mfp}).  We follow \citet{2013MNRAS.428.2565T}
by advecting the cold and hot populations passively with the
hydrodynamic flow, while allowing them to interact by a set of rate
equations governing the number densities of each population. The
planetary gas is initialized to be mostly neutral, cold hydrogen at the surface
of the planet, while the stellar wind gas is assumed to be so hot
that there is initially negligible hot neutral hydrogen atoms at the outer
boundary. Where the two fluids meet (the contact discontinuity),
charge exchange reactions may then create hot hydrogen, which traces
out the boundary between the two fluids.

\section{ Numerical Method }
\label{sec:numericalmethod}

The Zeus-MP hydrodynamics code \citep{2006ApJS..165..188H} is used
to solve for the mass density $\rho$, velocity $\bf v$, and internal
energy density $e$ using the hydrodynamic equations:

\begin{eqnarray}
\frac{\partial \rho}{\partial t} + {\bf \nabla} \cdot \left(\rho {\bf v}\right) & = & 0 \\
\frac{\partial (\rho {\bf v})}{\partial t} + {\bf \nabla}\cdot \left(\rho {\bf v v}\right) & = & -{\bf \nabla}p + \rho {\bf g} \\
\frac{\partial e}{\partial t} + {\bf \nabla}\cdot \left(e{\bf v}\right) & = & -p{\bf \nabla}\cdot{\bf v} \,\, .
\end{eqnarray}
A polytropic equation of state $e=p/(\gamma-1)$ is used, where $p$
is the pressure and the choice $\gamma = 1.01$ for the adiabatic
index allows the planetary and stellar wind gases to separately
remain roughly isothermal, although intermediate temperatures are
produced in the mixing layers between the two gases. The equations
are solved in two-dimensional $r-\theta$ spherical polar geometry assuming
axisymmetry. The stellar wind enters the grid from the $+\hat{z}$
direction. The inner radial boundary is taken to be the planet's
radius at $r = R_{\rm p}$. The outer boundary is chosen to be at
$r=50\, R_{\rm p}$. 

The gravitational acceleration ${\bf g}$ only includes the contribution of the planet,
\begin{equation}
{\bf g} = - \frac{G M_{\rm p}}{r^2} {\bf \hat{r}}
\end{equation}
\noindent where $M_{\rm p}$ is the mass of the planet and $G$ is
the gravitational constant.  \red{For simplicity, tidal gravity due to the host star has been omitted.  This would serve
to further accelerate the gas resulting in a faster outflow and increased ram pressure \citep{2009ApJ...693...23M}.}

The $\theta$ grid is equally spaced and extends from $0 \leq \theta
\leq \pi$. To resolve the wind launching region near the planet,
non-uniform spacing in $r$ is used, with $\delta r/r \approx 1.0075$.
For the fiducial resolution of $1024\times416$, this results in
zones with comparable radial and polar sizes, $\delta r_i \approx
r_i\delta \theta_i$.

The simulations discussed in this paper do not include self-consistent
launching of the wind by consideration of microphysical heating and
cooling processes. Rather, the assumption of a nearly isothermal
gas implicitly leads to an outflow since energy losses to adiabatic
expansion are overcome through adding in enough heat to keep the
gas isothermal. The goal of this paper is not to discuss the details
of how the planetary outflow is launched, but rather to conduct a
parametrized survey of the planetary wind and interaction with the
stellar wind as a function of the binding parameter $\lambda$, which
acts as a proxy for the amount of heating by the star. Presumably
the sequence of isothermal calculations could be mapped to more
detailed calculations \citep{2013Icar..226.1678K,2013Icar..226.1695K}
with similar effective $\lambda$.

The present calculation improves on that of \citet{2013MNRAS.428.2565T}
by using the correct geometry for the problem. \citet{2013MNRAS.428.2565T}
used two-dimensional Cartesian coordinates. An example of how their
calculation will differ from the present is in the density distribution of the
planetary wind with distance from the planet. For a constant wind
mass loss rate and wind velocity our calculation gives $\rho \propto
r^{-2}$ while theirs gives $\rho \propto R^{-1}$, where $R$ is the
polar radius in the x-y plane. Accurate determination of the mass
density possibly far from the planet then requires the correct
geometry.

\subsection{Ionization and Charge Exchange Reactions} 

The number densities of the hot and cold populations are computed
including four effects: advection with the hydrodynamic flow,
charge exchange reactions between hot and cold populations, 
optically-thin photoionization and radiative recombination reactions,  and collisional ionization by hot electrons. 
Photo- and collisional ionization are crucial
for determining the extent of the downstream neutral hydrogen
population. The treatment of charge exchange reactions follows that
in \citet{2013MNRAS.428.2565T}. Four species are defined: \red{neutral hot hydrogen and
hot protons} ($n_{\rm hot}^0$ and $n_{\rm hot}^+$,
respectively) and \red{neutral cold hydrogen and cold protons}  ($n_{\rm cold}^0$
and $n_{\rm cold}^+$, respectively).  The advection and reaction
steps are operator split, with the advection step accomplished by
the built-in capabilities of the Zeus code \citep{2006ApJS..165..188H}.
The reaction step solves the following rate equations for each
species:

\begin{eqnarray}
\label {eqn:tcchem1}
\frac{\partial n^0_{\rm hot}}{\partial t} & = & \beta\left(n^+_{\rm hot} n^0_{\rm cold} - n^0_{\rm hot} n^+_{\rm cold}\right) \nonumber \\
 & & - \Gamma n^0_{\rm hot}  + \alpha_{\rm hot}(n^+_{\rm hot})^2 - C_{\rm hot}n^0_{\rm hot}n^+_{\rm hot}\\
\label {eqn:tcchem2}
\frac{\partial n^0_{\rm cold}}{\partial t} & = &  -\beta\left(n^+_{\rm hot} n^0_{\rm cold} - n^0_{\rm hot} n^+_{\rm cold}\right) \nonumber \\
 & & - \Gamma n^0_{\rm cold}  + \alpha_{\rm cold}(n^+_{\rm cold})^2 - C_{\rm hot}n^0_{\rm cold}n^+_{\rm hot}\\
\label {eqn:tcchem3}
\frac{\partial n^+_{\rm hot}}{\partial t} & = & - \frac{\partial n^0_{\rm hot}}{\partial t} \\
\label {eqn:tcchem4}
\frac{\partial n^+_{\rm cold}}{\partial t} & = & -\frac{\partial n^0_{\rm cold}}{\partial t}
\end{eqnarray}

\noindent where $\alpha_{\rm hot}$ and $\alpha_{\rm cold}$ are the recombination rates for
the hot and cold species, respectively.  The collisional ionization rate by hot electrons, $C_{\rm hot}$, is taken to be $2.74\times 10^{-8}\,{\rm cm^{3}\, s^{-1}}$ \citep{janev2003collision}.  Since the thermal energy of the cold electrons is much less than the ionization potential of hydrogen, ionization by cold electrons can safely be ignored\footnote{For comparison, the collisional ionization rate by electrons with $T=10^4\ {\rm K}$ 
is $6.27\times 10^{-16}\,{\rm cm^{3}\,s^{-1}}$.}. The charge-exchange rate is taken to
be $\beta = 4\times 10^{-8}\,{\rm cm^3\, s^{-1}}$ and is assumed
to be constant.  Separate recombination rates are employed for the
hot and cold populations due to the differing characteristic
temperatures and densities. Photoionization of hydrogen is given
by the rate $\Gamma = (6.9\ {\rm hrs})^{-1}$ \citep{2011ApJ...728..152T}.  The
optically thin rate is used in the entire simulation volume for
simplicity.  This will overestimate the ionization rate at large
optical depths, especially in the region in the planet's shadow,
and hence underestimate transit depths. For the hot species, existing
primarily in the diffuse, ionized stellar wind, the Case A recombination
rate is used with a characteristic temperature of $T = 10^6\,{\rm
K}$, giving a rate coefficient $\alpha_{\rm hot} = 9.677\times
10^{-15}\, {\rm cm^3\, s^{-1}}$.  Due to the higher density and
larger neutral fraction of the planetary wind, the Case B value for
the cold recombination rate, resulting in $\alpha_{\rm cold } =
2.59\times 10^{-13}\,{\rm cm^3\, s^{-1}}$ \citep{2011piim.book.....D}.
For simplicity, recombination between hot protons and cold electrons,
and cold protons and hot electrons has been ignored. This will have
the effect of underestimating the number of neutral hydrogen atoms.
Equations \ref{eqn:tcchem1}-\ref{eqn:tcchem4} are discretized using
backward differencing for stability, and the resulting system of
equations are solved using Newton-Raphson iteration.

The advection-reaction steps used to determine the number densities
$n_{\rm hot}^+$, $n_{\rm hot}^0$, $n_{\rm cold}^+$ and $n_{\rm
cold}^0$ do not affect the basic hydrodynamic flow variables $\rho$,
$\bf v$ and $e$. The fluid velocity $\bf v$ is used, however, 
for the advection step, and the number densities are related back to $\rho$ by
\begin{equation}
\rho  =  m_{\rm p} \left( n^0_{\rm cold} + n^+_{\rm cold} + n^0_{\rm hot} + n^+_{\rm hot}  \right)\,\, .
\end{equation}

\subsection{Boundary Conditions}

\begin{deluxetable}{lr}
\tablecolumns{2}
\tablewidth{0pc}
\tablecaption{HD 209458b Parameters}
\tablehead{\colhead{} & \colhead {}}
\startdata
\sidehead{\em Planet Parameters}
Mass$^a$ $M_{\rm p}$($M_{\rm J}$) & $0.714$\\
Radius$^a$ $R_{\rm p}$($R_{\rm J}$) & $1.46$ \\
$\rho_{\rm p,b}$ ($\rm g\, cm^{-3}$) & $10^8\, {\rm m_{\rm H}}$ \\
\sidehead{\em Star Parameters}
Radius$^a$ $R_\star$ ($R_\odot$) & $1.2$ \\
\sidehead{\em Stellar Wind Parameters}
$n_\star$ ($\rm cm^{-3}$) & $10^4$ \\
$T_\star$ ($\rm K$)& $10^6$ \\
$v_\star$ ($\rm km\, s^{-1}$) & 200 \\
\sidehead{\em Reaction Rates}
$\Gamma$ ($\rm s^{-1}$) & $4\times 10^{-5}$ \\
$\alpha_{\rm cold}$ ($\rm cm^{3}\, s^{-1}$) & $2.59\times 10^{-13}$ \\
$\alpha_{\rm hot}$ ($\rm cm^{3}\, s^{-1}$)& $9.677\times 10^{-15}$ \\
$\beta$ ($\rm cm^{3}\, s^{-1}$)& $4\times 10^{-8}$ \\
$C_{\rm hot}\, ({\rm cm^3\, s^{-1}})$ & $2.74\times 10^{-8}$\\
\enddata
\tablenotetext{a}{Source: exoplanet.eu}
\label{table:209458b}
\end{deluxetable}

The inner boundary condition at the surface of the planet is set
to be reflective in all hydrodynamic variables. The density $\rho_{\rm p,b}$ at the base (see Table \ref{table:209458b})
is characteristic of the layer where stellar EUV radiation is absorbed, 
which ionizes and heats the gas to a temperature $T_{\rm p} \sim 10^4\ {\rm K}$
(see e.g. \citealt{2011ApJ...728..152T}).
To launch the wind,
the density in the first active zone is fixed at $\rho_{\rm p,b}$.
While the inner boundary conditions prevent gas from entering or
leaving through the boundary, the first active zones in the radial
direction become sources for gas, regulating the outflow. The cold
species are set to their photoionization equilibrium values resulting
in an ionization fraction $n_{\rm p,cold}/(n_{\rm H,cold}+n_{\rm
p,cold}) \approx 0.7$, and  there are no hot species at the base
of the planetary atmosphere.  The energy in the first active zone
is fixed based on the escape parameter $\lambda$,
\begin{equation}
e(R_{\rm p},\theta) = \frac{G M_{\rm p}\rho_{\rm p,b}}{\left(\gamma-1\right)R_{\rm p}\lambda}\,\, .
\end{equation}

We first let the spherically-symmetric planetary wind develop freely 
in the absence of a stellar wind, by imposing an outflow boundary condition 
at $r=R_{\rm outer}$.  After the planetary wind reaches a steady state, typically
after a few $\times 10^5\, {\rm s}$, the stellar wind is turned on.
This is done by imposing an inflow boundary condition for the half 
of outer radial boundary with $\theta < \pi/2$, while the other half 
(with $\pi/2 < \theta < \pi$) retains the outflow boundary condition.
The inflow velocity is set to be parallel to the star-planet axis with a
constant initial velocity $v_{\star}$. Specifically,

\begin{eqnarray}
v_r\left(R_{\rm outer},\theta\right) & = & - v_{\star}\cos\theta \\
v_\theta\left(R_{\rm outer},\theta\right) & = & v_{\star}\sin\theta\,\, .
\end{eqnarray}

Along the inflow boundary, the gas is taken to be purely \red{hot protons and electrons}
 with energy $e_\star=2n_\star k T_\star/(\gamma-1)$.  The
assumed values for the stellar wind parameters $n_\star$ and $T_\star$
are given in Table \ref{table:209458b}.

\section{Structure of the Gas}
\label{sec:hotcolddist}

\begin{figure*}
\epsscale{1.2}
\plotone{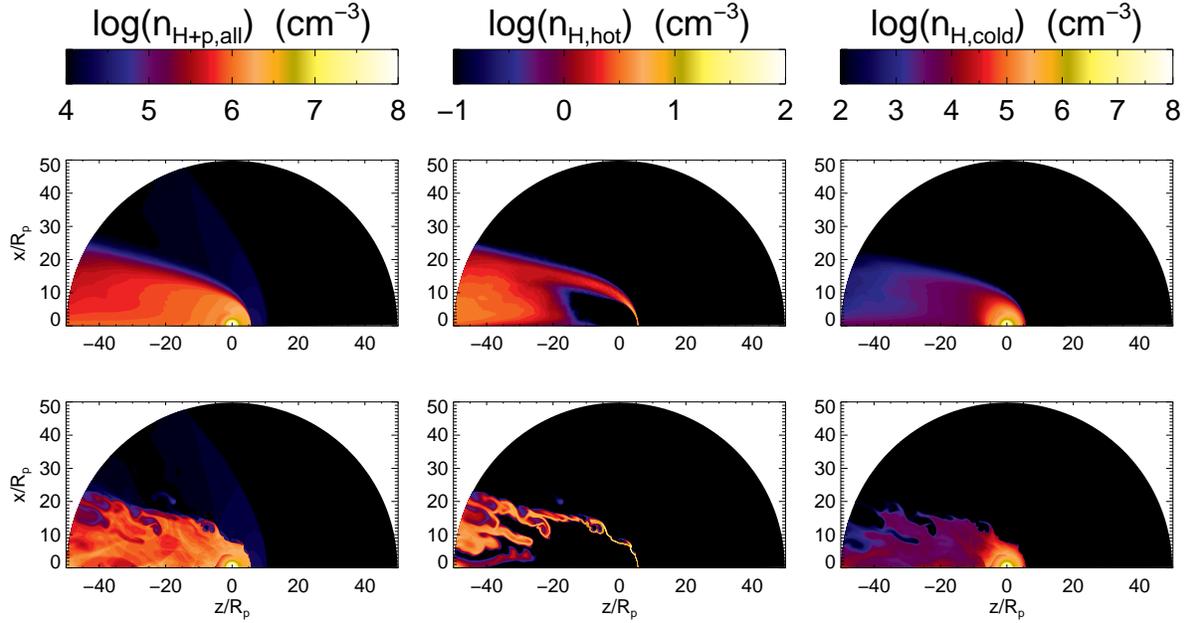}
\caption{ Contour plots of particle densities for $\lambda=4.5$ as a function of position.
Position is in units of the planetary radius, $R_p$.
The upstream direction is to the right in each plot,
and downstream is to the left. The upper three panels are time-averaged values
while instantaneous values at the end of the run are in the bottom row.
The two left panels show the total density $n_{\rm H+p, all}=\rho/m_p$, including hot and cold
neutrals and protons. The two center panels
show the density of hot hydrogen, and the two right panels show cold hydrogen.
}
\label{Fig:InstVsAvg}
\end{figure*}

In this section, numerical results are presented for the populations
of cold and hot hydrogen, as derived from Zeus simulations. The
goal is to perform a parameter study in the escape parameter,
$\lambda$, and to understand how the distribution changes for each
population. \red{Values of $\lambda$ between 3.5 and 8 have been investigated as this range spans both the colliding winds and viscous regimes.}These results will then be used in \S \ref{sec:lymanalpha}
to understand the Lyman $\alpha$ transmission spectrum.

Figure \ref{Fig:InstVsAvg} shows the distribution of the total
particle density $\rho/m_p$, as well as that of the hot and cold
populations of neutrals \red{for $\lambda=4.5$ run, a case where the planetary outflow
approaches the sound speed but never becomes supersonic.} Comparing the time averaged values in the
three upper panels to the instantaneous values in the three lower
panels shows that the regions out near the mixing layer show
significant variability. This is due to the Kelvin-Helmholtz
instability between the slowly moving planetary gas and the fast
stellar wind \citep{2013MNRAS.428.2565T}. The variability is larger for the hot neutrals
than cold since the former arise explicitly in the mixing layer,
while the latter are the extension of the planet's atmosphere, which
is assumed static at the base of the grid.  Unless otherwise
specified, we time average our hydrodynamic quantities from $t=5\times
10^5\, {\rm s}$ to $t=10^6\, {\rm s}$, the end of the run. This
serves to smooth out transient turbulent structures within the flow.

Comparison of the left and right panels in Figure \ref{Fig:InstVsAvg}
shows that the gas is predominantly neutral at the inner radial
boundary, and the cold population is highly ionized out beyond
$r \sim 5\ R_p$. The cold population is strongly peaked toward the
planet. By contrast, the middle panels show that the hot population
forms a thin sheath in the mixing layer (for this value of $\lambda$),
well outside the bulk of the cold population,
with negligible values interior or exterior to the sheath. Note the
difference in density scale in the middle panels. The optical depth for Lyman $\alpha$ absorption is the
product of number density, path length and absorption cross section. While the
cross section is $\sim 10^5$ larger for the hot population, the
particle densities are orders of magnitude smaller, and, unless
looking along a sightline through the sheath, the length over which
particle densities are significant is much smaller than that for the thermal
distribution. Careful numerical integration is required to accurately
compute the optical depths of the hot population, as done in \S\ref{sec:lymanalpha}. 

\begin{figure*}
\epsscale{1.2}
\plotone{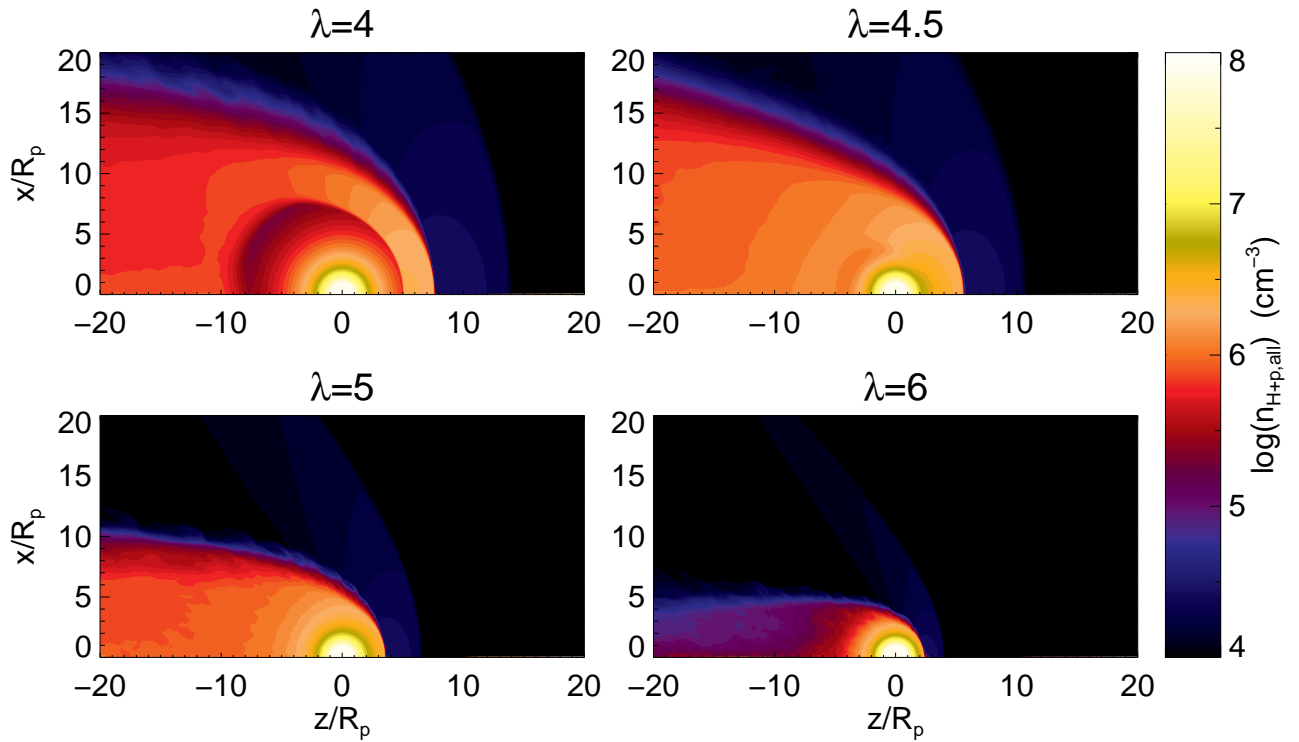}
\caption{Contours of time-averaged total number density, $n_{\rm H+p,all}=\rho/m_p$, 
for $\lambda=4,4.5,5,6$.}
\label{Fig:DensContour}
\end{figure*}

\begin{figure*}
\epsscale{1.0}
\plotone{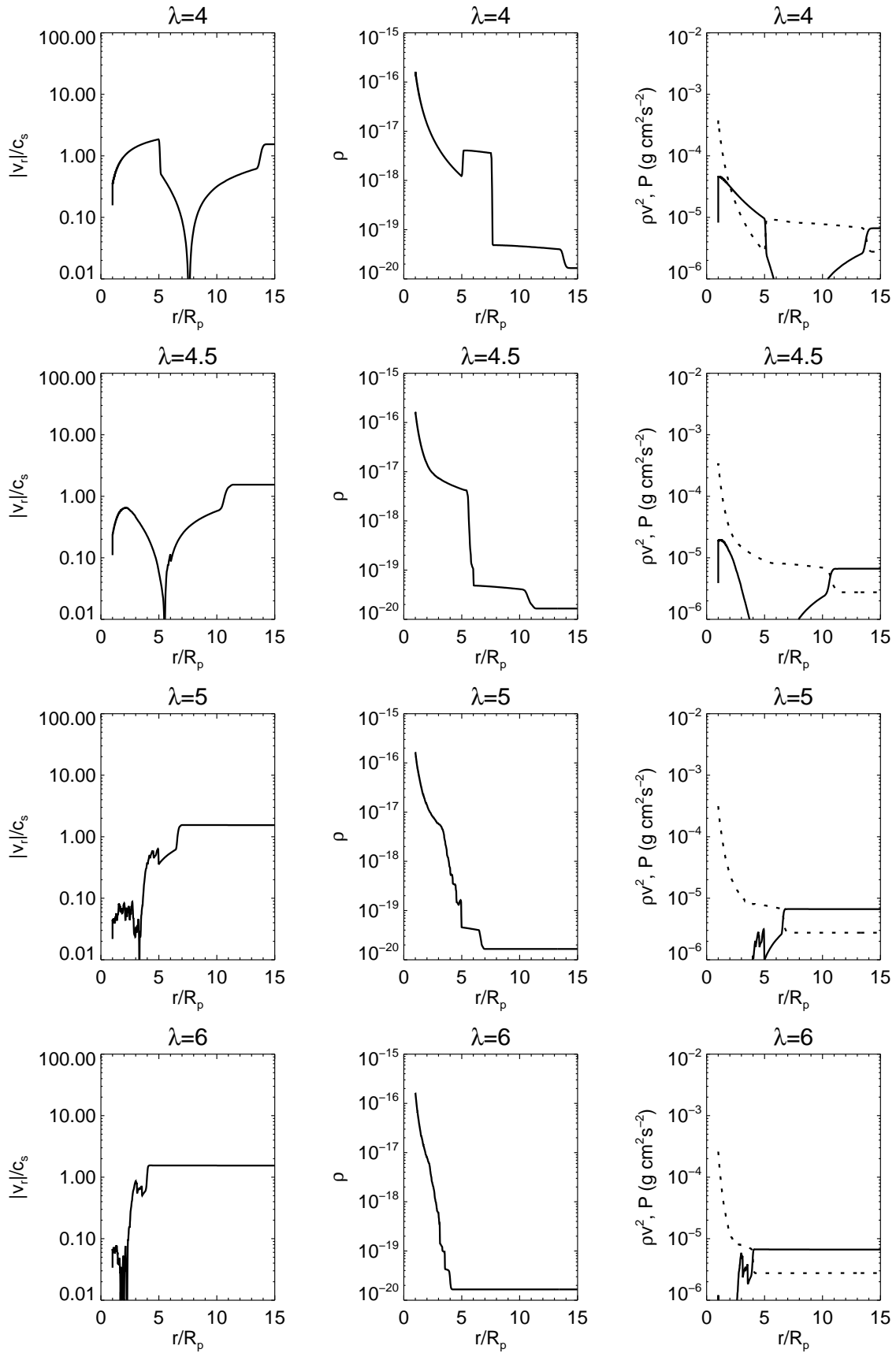}
\caption{ Fluid variables versus radius along the substellar line toward the star.
Left Column: The local Mach number $|v_r|/c_s$ where $c_s$ is the local sound speed. Center Column: density in units of ${\rm g\ cm^{-3}}$.
Right Column: Ram pressure ($\rho v^2$, solid line) and thermal pressure ($P$, dotted line).}
\label{Fig:VelocityPlots}
\end{figure*}
\subsection{Mass Density}
Figure \ref{Fig:DensContour} shows the total mass density for the
simulations and Figure \ref{Fig:VelocityPlots} shows one-dimensional
plots of Mach number, density and pressure along the substellar
line.  The two extreme cases are $\lambda=4$ (colliding winds) and
$\lambda=6$ (viscous). The other two cases, $\lambda=4.5$ and $5$,
show the transition between the two regimes.

\begin{figure*}
\epsscale{1.2}
\plotone{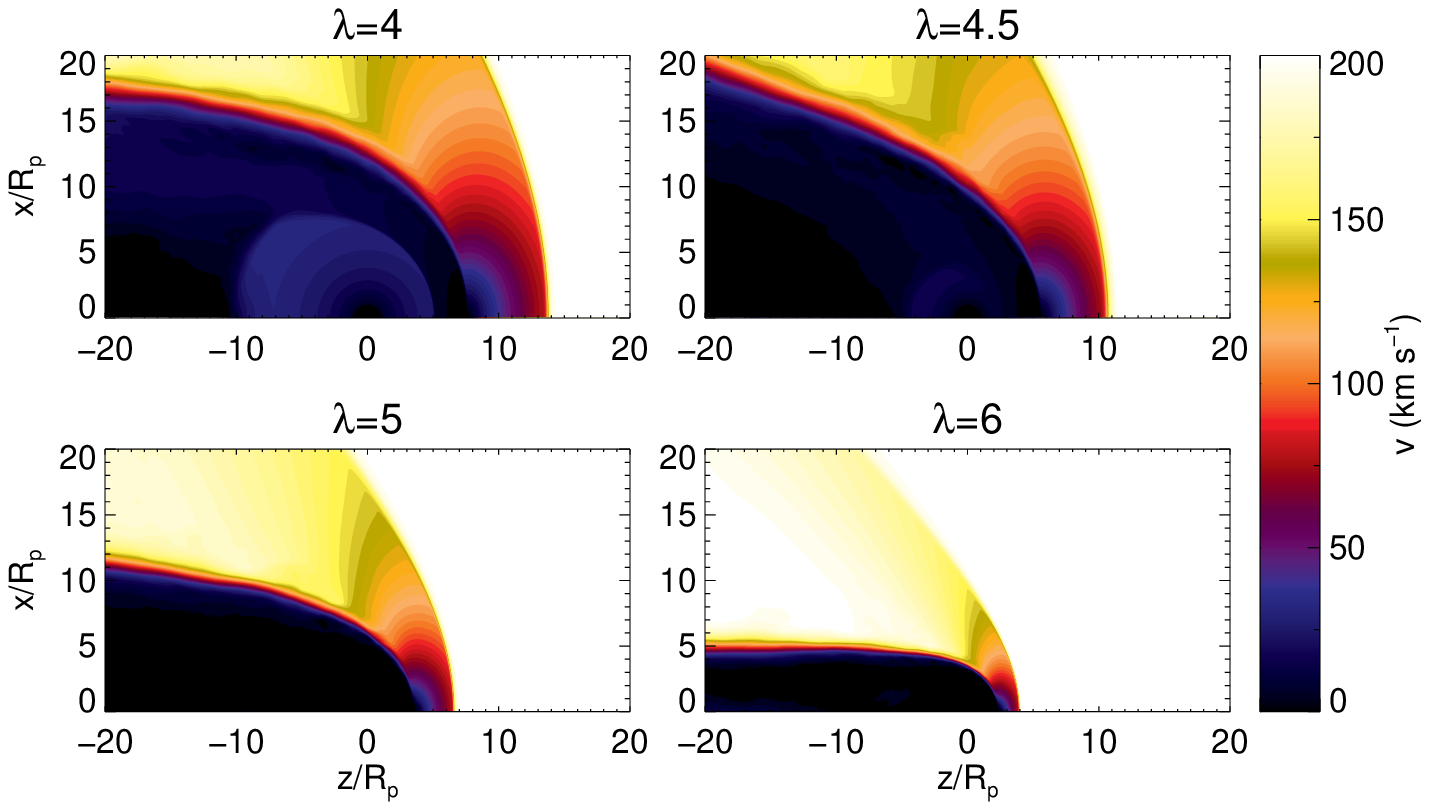}
\caption{Contour plots of time-averaged speed for $\lambda = 4, 4.5,
  5, 6$. }
\label{Fig:SpeedContour}
\end{figure*}

The $\lambda=4$ case is shown in the top left panel of Figure
\ref{Fig:DensContour} and the top row in Figure \ref{Fig:VelocityPlots}.
The density profile is nearly hydrostatic near the planet where the
velocity is subsonic. The planetary gas accelerates to the sound
speed at $r_s=(\lambda/2)R_p=2R_p$. At $r \simeq 5R_p$, a shock
decelerates the planetary flow, leading to a high-density post-shock
shell. The stellar wind also experiences a deceleration (bow) shock
(at $r=14\ R_p$, see Figure \ref{Fig:SpeedContour} ) and density increase in the post-shock flow,
although this occurs at such low densities it is hard to see in
Figure \ref{Fig:DensContour}.  For the isothermal flow considered
here, the density jump at a shock goes as $(v/c)^2$, the upstream
Mach number.  The velocity $v_r \rightarrow 0$ for both planetary
and stellar wind gases at $r \simeq 7.5\ R_p$, the contact
discontinuity.  There is thermal pressure balance at this point,
since the ram pressure is zero.

In this $\lambda=4$ colliding-winds case (top left panel in
Figure \ref{Fig:DensContour} and the top row in Figure
\ref{Fig:VelocityPlots}, the mass loss rate $\dot{M}$ from
the planet is due to acceleration of the flow by pressure gradients
near the surface of the planet, and is independent of the details
of the colliding planetary and stellar winds. An important role of
the stellar wind is to turn the planetary wind initially heading
toward the star around, so that it eventually forms a tail behind
the planet.  Ignoring mixing for the moment, the tail will asymptote
to a cylindrical radius $r_t$ set by thermal pressure balance
$\rho_{p,t} c_p^2 = \rho_\star c_\star^2$, where $\rho_{p,t}$ is
the mass density in the tail and $\rho_\star = m_p n_\star$ is the
mass density in the stellar wind.  For speed $v_{p,t}$ of the gas
in the tail, mass continuity implies
\begin{eqnarray}
\dot{M} & = & \pi r_t^2\ v_{p,t} \rho_{p,t} = 
\pi r_t^2\ v_{p,t} \rho_\star \left( \frac{c_\star}{c_p} \right)^2.
\label{eq:tail}
\end{eqnarray}
While $\dot{M}$ and $\rho_{p,t}$ are immediately known, the solution
for $v_{p,t}$ is more complicated, as it involves solving the
Bernoulli equation in the post-shock flow, as it bends around to
form the tail. However, if $v_{p,t} \simeq {\rm (a\ few)} \times
c_p$ is treated as given, since $\dot{M}$, $v_{p,t}$ and $\rho_{p,t}$
are known, then equation \ref{eq:tail} can be used to solve for the
radius $r_t$ of the tail region. Figure \ref{Fig:DensContour} shows
that for $\lambda=4$, $r_t \simeq 15\ R_p$, and $\rho_{p,t}/m_p
\simeq 10^6\ {\rm cm^{-3}} \simeq 10^2 n_\star \simeq (c_\star/c_p)^2
n_\star$, as expected for thermal pressure balance.  Hence the
radius $r_t$ is set by allowing the density of the planetary gas
to decrease until this pressure balance is achieved. Lastly, since
$v_{p,t} \neq v_\star$ when the tail first forms near the planet,
the shear between planetary and stellar wind gases will lead to a
mixing layer extending from the edges to the center of the tail.
In this colliding winds case, the tail is only well mixed at many
tens of $R_p$ behind the planet (and is not apparent in Figure
\ref{Fig:DensContour}).

For the $\lambda = 6$ case in Figure \ref{Fig:DensContour}, the
stellar wind bow shock occurs near $r=5R_p$, much closer to
the planet than the $\lambda=4$ case, and the post-shock stellar
wind gas extends closer to the planet, inside of where the isothermal
sonic point radius should have been, at $r_s=3R_p$. The planet
exhibits a narrow tail with $r_t \sim {\rm (a\ few)} \times R_p$,
and as will become apparent in the following plots, the tail is
well mixed even immediately behind the planet. In this case, the
planetary gas is very subsonic out to the mixing layer, and $\dot{M}$
is set by the frictional coupling between the two gases.

The $\lambda=4.5$ and $5$ cases in Figure \ref{Fig:DensContour} and
the middle rows of Figure \ref{Fig:VelocityPlots} are more complicated,
as the planetary gas approaches, but does not exceed the sound
speed. The tail radius $r_t$ decreases steadily from $\lambda=4$
to $\lambda=6$.

\begin{figure*}
\epsscale{1.2}
\plotone{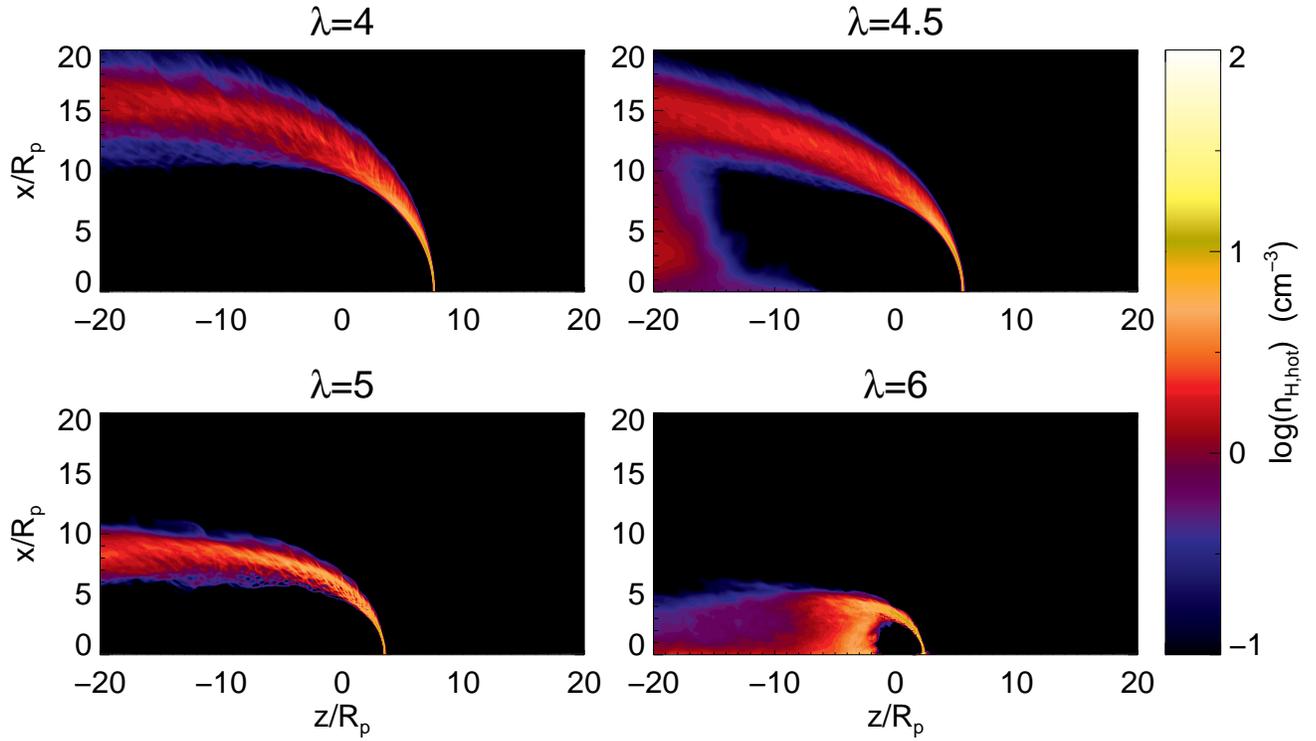}
\caption{Contour plots of time-averaged number density of the hot hydrogen population
for $\lambda = 4, 4.5, 5, 6$. }
\label{Fig:HotHContour}
\end{figure*}

\begin{figure*}
\epsscale{1.2}
\plotone{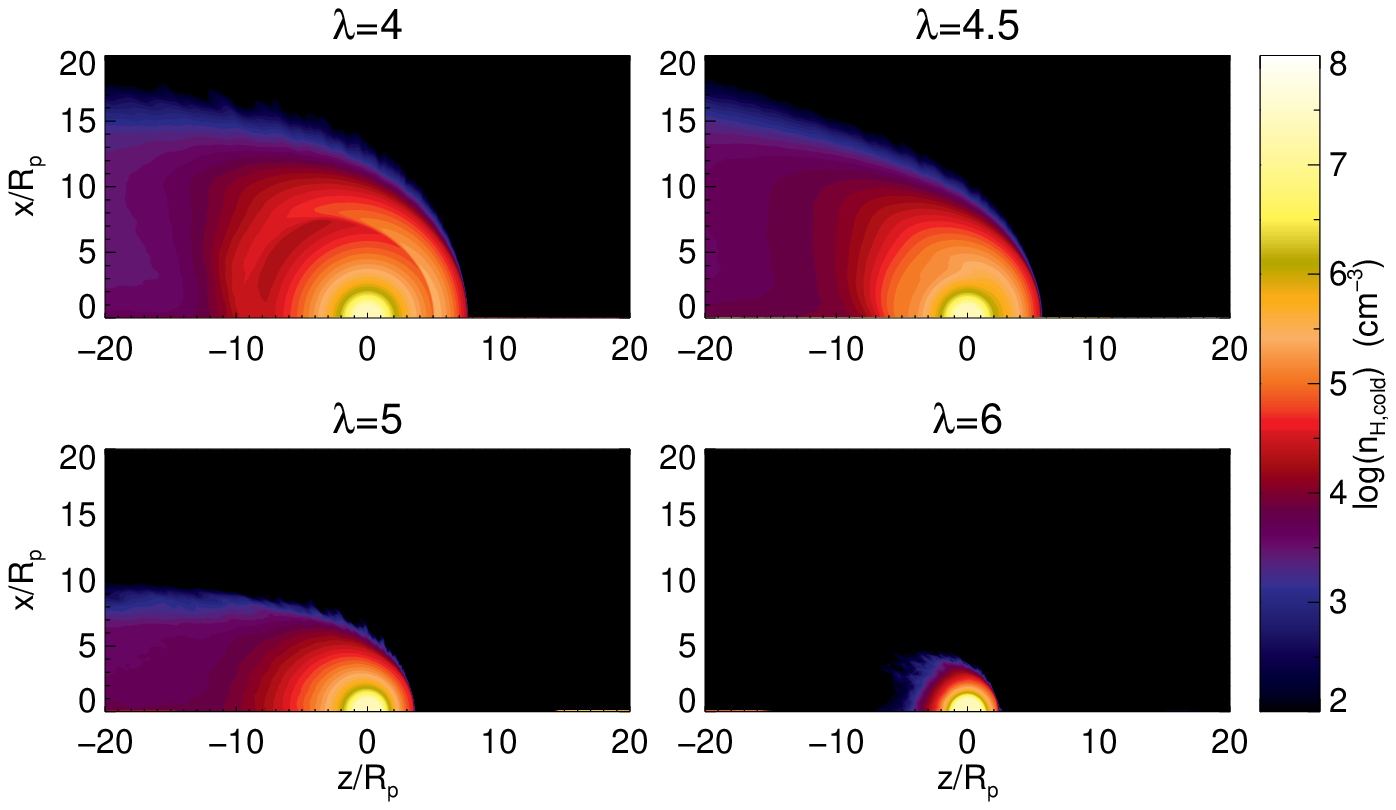}
\caption{Contour plots of time-averaged number density of the cold hydrogen population 
for $\lambda = 4, 4.5, 5, 6$.  }
\label{Fig:ColdHContour}
\end{figure*}

\subsection{Hot and Cold Populations}
The distributions of the cold and hot hydrogen populations are shown
in Figures \ref{Fig:HotHContour} and \ref{Fig:ColdHContour} for the
same values of $\lambda$ as in Figures \ref{Fig:DensContour} and
\ref{Fig:VelocityPlots}. The hot hydrogen is created near
the contact discontinuity, where the planetary and
stellar wind mass densities in the mixing layer are related by
$\rho_{p} = \rho_\star (c_\star/c_p)^2 \simeq 100 \rho_\star$, due
to thermal pressure balance. If charge exchange reactions dominate
photoionization and recombination, and are in equilibrium there,
then $n^0_{\rm hot} \simeq n^0_{\rm cold}(n^+_{\rm hot}/ n^+_{\rm
cold}) = n^+_{\rm cold} (n^0_{\rm cold}/n^+_{\rm cold})(n^+_{\rm
hot}/ n^+_{\rm cold}) \simeq 10^{-3} n^+_{\rm cold}$, since the
ionization level in that case is $n^0_{\rm cold}/n^+_{\rm cold}
\sim 0.1$. The end result is that, at the contact discontinuity, the hot hydrogen density is several orders
of magnitude smaller than the total particle density in the neighboring
planetary gas. Additionally, the planetary gas at the contact discontinuity
is smaller than that at the base, where stellar EUV is absorbed and the gas
is heated and ionized,  by another 3-4 orders of magnitude. For the same base density,
Figure \ref{Fig:VelocityPlots} shows that the planetary gas density at the contact discontinuity
can vary by order an order of magnitude from the $\lambda=4$ to the $\lambda=6$ cases.
Hence the source of hydrogen atoms for charge exchange is sensitive to the boundary conditions
much deeper in the planet's atmosphere.

The population of hot hydrogen requires cold hydrogen and hot protons
present to interact. There is a remarkable difference in the
morphology as $\lambda$ is increased.  In the colliding winds
($\lambda=4$) case, the number density of hot hydrogen is largest
near the substellar mixing layer. However, that layer is very thin.
There is a sheath of hot hydrogen a factor of 10 lower density, but
extending over a region many 10's times larger than the substellar region.
Hence the sheath surrounding the tail may be equally important
as the substellar region for a favorable viewing geometry through
the sheath. Also notice that there is little mixing into the central
region (around the $z$ axis) of the tail for the $\lambda=4$ case.  This is unlike the $\lambda=6$
case, where hot hydrogen is well mixed into the tail region.  Although
the density on the substellar side is again a factor of 10 higher, 
the path length through the tail is much larger, and the tail may
be of comparable importance in the optical depth.  The $\lambda=4.5$
and $5$ cases illustrate the transition between the two extreme cases, showing
the sheath size decreasing, and hot hydrogen eventually being mixed
into the tail as $\lambda$ increases.

The cold hydrogen population in Figure \ref{Fig:ColdHContour} is
always strongly concentrated around the planet. The high density,
post-shock gas is clearly visible for the colliding winds case for
$\lambda=4$. The size of the cold hydrogen distribution decreases
as $\lambda$ increases, and is sharply cut off by the stellar wind
for $\lambda=5$ and $6$. Note the difference in color scale for the
hot hydrogen in Figure \ref{Fig:HotHContour} and the cold hydrogen
in Figure \ref{Fig:ColdHContour}.

Figure \ref{Fig:SpeedContour} shows contour plots of $|{\bf v}|$ as a function of position. Immediately apparent is the deceleration of the stellar wind near the planet, and subsequent re-acceleration downstream. A shear layer between planetary and stellar wind gas is also easily seen in the plots. Only the $\lambda=4$ case shows a clear acceleration of gas near the planet, while other cases have velocities $|{\bf v}| \la 20-30\ {\rm km\ s^{-1}}$.

\section{ mass loss rate }
\label{sec:mlr}
The goal of this section is to discuss the steady state mass loss
rate $\dot{M}$ from the planet as a function of $\lambda$ and stellar wind parameters.   \red{The mass loss rates are estimated analytically and compared to the results of the simulations.    Planetary magnetic fields, not considered here, can serve to trap planetary gas and limit mass loss rates (see \citealt{2011ApJ...728..152T,2014ApJ...788..161T,2013MNRAS.428.2565T,2014MNRAS.444.3761O} for a more complete discussion).}

\begin{figure}
\epsscale{1.2}
\plotone{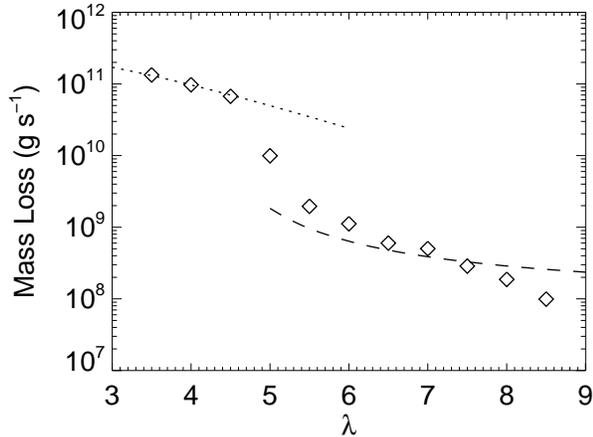}
\caption{Mass-loss rate versus $\lambda$ for our parameter study. Diamonds indicate mass loss rates for specific simulations.  
The dotted line is the mass-loss rate estimated analytically for a supersonic 
wind as in eq. \ref{eqn:masslosswind}.  The dashed line is the analytic estimate for
  an atmosphere being stripped by the stellar wind, as in eq. \ref{eqn:masslossstripped}}
\label{Fig:MassLossRate}
\end{figure}

The mass loss rate has been computed for the $\lambda=3.5-8.5$
by an integral over a sphere near the inner boundary:
\begin{equation}
\dot{M} = 2\pi R_{\rm p}^2\int_0^\pi  \rho\left(R_{\rm p}+\epsilon,\theta\right)v_{r}\left(R_{\rm p}+\epsilon,\theta\right)\sin\theta d\theta.
\label{eq:mdot}
\end{equation}
Here the radius $R_{\rm p}+\epsilon$ is meant to represent a position
slightly outside the planet's surface, and in practice we compute
the integral in eq.\ref{eq:mdot} at the 4th active zone in radius.

Numerical results for the mass loss rate versus $\lambda$ are
presented in Figure \ref{Fig:MassLossRate}.  As $\lambda$ increases, there is initially an
exponential decrease $\dot{M} \propto e^{-\lambda}$, followed by a transition
zone in the region $\lambda \sim 5-6$ with steeper slope, and then a
much slower decrease for $\lambda \gtrsim 6$. Analytic estimates are 
derived below for the small and large $\lambda$ limits, while the
intermediate case, showing rapid decrease in $\dot{M}$ is harder
to understand analytically as flow speeds $v_r \sim c_p$ and the
density and velocity profiles are complicated.

In the small $\lambda$ and \red{relatively }weak stellar wind case, the planetary
wind becomes supersonic and hence the mass loss rate is given by
the well-known isothermal formula, which is now briefly reviewed.
Since $\dot{M} = 4\pi r^2 \rho v_r$, it is possible to calculate
$\dot{M}$ at the sonic point where $r=(\lambda/2)R_{\rm p}$
(eq.\ref{eq:rsonic}), $v_r=c_p$ and the sonic point density can
be related to the density at the base of the outflow using Bernoulli's
equation as
\begin{equation}
\rho_{\rm p,s}\approx \rho_{\rm p,b}\exp\left(\frac{3}{2}-\lambda\right)\,\, .
\label{eqn:rhosonic}
\end{equation}
The standard result is
\begin{equation}
\dot{M} \approx \pi \rho_{\rm p,b} \left(GM_{\rm p} R_{\rm p}^3\lambda^3\right)^{1/2}\exp\left(\frac{3}{2}-\lambda\right)\,\, .
\label{eqn:masslosswind}
\end{equation}
Note that this expression depends only on planetary parameters and
is independent of the stellar wind.  The value of $\lambda$, through
$T_{\rm p}$, is strongly dependent on the stellar radiation though.
For the case of HD 209458b, the mass loss was estimated by
\citet{2013Icar..226.1678K} to be in the range $\dot{M} = 4-6 \times
10^{10}\,{\rm g\, s^{-1}}$.   Using eq. \ref{eqn:masslosswind} and
the assumed base density $\rho_{\rm p,b}=10^8\ m_{\rm H}$, this corresponds
to $\lambda = 4.7-5.3$, values that, for the parameters in this paper, correspond to the transition 
between the viscous and colliding wind regimes. There is excellent agreement between
eq.\ref{eqn:masslosswind} and the numerical results in Figure
\ref{Fig:MassLossRate} for small $\lambda$.

In the opposite limit of large $\lambda$ and a \red{relatively} strong stellar wind,
the planetary wind is highly subsonic with a nearly hydrostatic
density profile out to the mixing layer. Mass loss is then not due
to the launching of a supersonic planetary wind, but instead the
viscous erosion of the atmosphere by the stellar wind. To estimate
the mass-loss rate, we use the turbulent regime mass loss rate
derived by \citet{1982MNRAS.198.1007N}, which we now briefly review. As the stellar wind
moves around the planetary gas, it is assumed that turbulence
develops near the boundary of the two fluids, driven e.g. by the
Kelvin-Helmholtz instability. Momentum is then transported from the
stellar wind fluid to the planetary gas by eddies. Assuming eddy
velocities $\sim v_\star$, the stress acting to accelerate the
planetary gas at the boundary scales as $\sim \rho_\star v_\star^2$.
Acting over an effective area $\sim \pi R_{\rm eff}^2$, the momentum
input to the planetary gas per unit time is $\sim \pi R_{\rm eff}^2
\rho_\star v_\star^2$. Assuming the planetary gas is accelerated
to a speed $\sim v_\star$, the mass loss rate is then
\begin{equation}
\dot{M}_{\rm turb} = \pi R_{\rm eff}^2 \rho_{\star} v_{\star}\,\, .
\label{eq:Mdotturb}
\end{equation}
The effective radius $R_{\rm eff}$ is estimated as the position
where thermal pressure from the hydrostatic atmosphere balances
stellar wind ram pressure:
\begin{eqnarray}
\frac{GM_{\rm p}\rho_{\rm p,b}}{R_{\rm p}\lambda}\exp\left[R_{\rm p}\lambda\left(\frac{1}{r}-\frac{1}{R_{\rm p}}\right)\right]
& =& \rho_{\star} v_\star^2.
\label{eq:balance}
\end{eqnarray}
The solution for $r=R_{\rm eff}$ is
\begin{equation}
\frac{R_{\rm eff}}{R_{\rm p}} = \frac{\lambda}{\log\left(\frac{v_\star^2\rho_\star R_{\rm p}\lambda}{GM_{\rm p}\rho_{\rm p,b}}\right) + \lambda}.
\label{eq:Reff}
\end{equation}
In the present simulation, the ratio of escape speed to stellar
wind speed $GM_{\rm p}/R_{\rm p}v_\star^2 \sim 10^{-2}$, while the
ratio of densities $\rho_{\rm p,b}/\rho_\star \sim 10^4$, so the
argument of the logarithm in eq.\ref{eq:Reff} is much smaller than
unity, and hence the denominator is less than $\lambda$, leading
to an effective radius larger than the planetary radius.  Plugging
eq. \ref{eq:Reff} into eq. \ref{eq:Mdotturb} yields a mass-loss rate
of
\begin{equation}
\dot{M}_{\rm turb} = \frac{\pi \lambda^2 R_{\rm p}^2\rho_{\star}v_{\star} }{\left(\log\left(\frac{v_\star^2\rho_\star R_{\rm p}\lambda}{GM_{\rm p}\rho_{\rm p,b}}\right) + \lambda\right)^2}
\label{eqn:masslossstripped}
\end{equation}
There is approximate agreement between eq. \ref{eqn:masslossstripped}
and the numerical results for large $\lambda$ in Figure
\ref{Fig:MassLossRate}. Since we have not performed detailed estimates
of turbulent viscosity at the boundary, or taken into account the
non-spherical geometry, eq.\ref{eqn:masslossstripped} gives
surprisingly good agreement with the numerical results.

A supersonic wind from the planet is only possible for a sufficiently weak stellar wind. The critical value of $\lambda$ ($\lambda_{\rm c}$)
separating the two regimes can be derived 
from eq. \ref{eq:balance} by setting $r=r_{\rm sonic}$, that is, 
by allowing $R_{\rm eff}$ to extend out to the sonic point. This
gives an equation,
\begin{eqnarray}
\lambda_{\rm c} e^{\lambda_{\rm c}} & =& e^2 \frac{GM_{\rm p}\rho_{\rm p,b}}
{v_\star^2\rho_\star R_{\rm p}}
\label{Eqn:lambdac}
\end{eqnarray}
to solve for $\lambda_{\rm c}$. For the parameters in table \ref{table:209458b} for HD 209458b,
we find $\lambda_{\rm c} \simeq 5.3$, which is near the value of $\lambda$ marking the transition
from the viscous regime and the colliding winds regime.

\begin{figure}
\epsscale{1.2}
\plotone{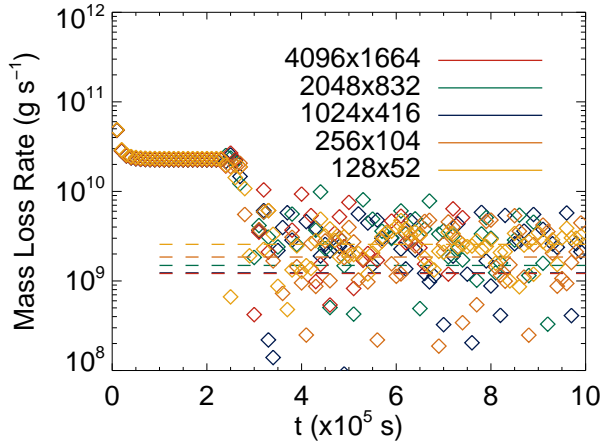}
\caption{The mass loss rate as a  function of time for $\lambda=6$ at various resolutions. The points are the instantaneous values, and the dashed lines are the time averages. The different colors represent different grid resolution, labelled by the number of radial and $\theta$ grid points, respectively.}
\label{Fig:MassLossResStudy}
\end{figure}

In the large $\lambda$, viscous regime, the mass loss rate is small, and fluctuates from positive to negative values depending on pressure fluctuations at the mixing layer. Due to the large variations in the instantaneous rate, it was found that long time averages were necessary to find the mean mass loss rate. Figure \ref{Fig:MassLossResStudy} shows $\dot{M}$ versus time for five different grid resolutions. It is found that as the number of grid points is increased, the time-averaged values of $\dot{M}$ converge.

\section{Mean Free Paths}
\label{sec:mfp}
\begin{figure}
\epsscale{1.2}
\plotone{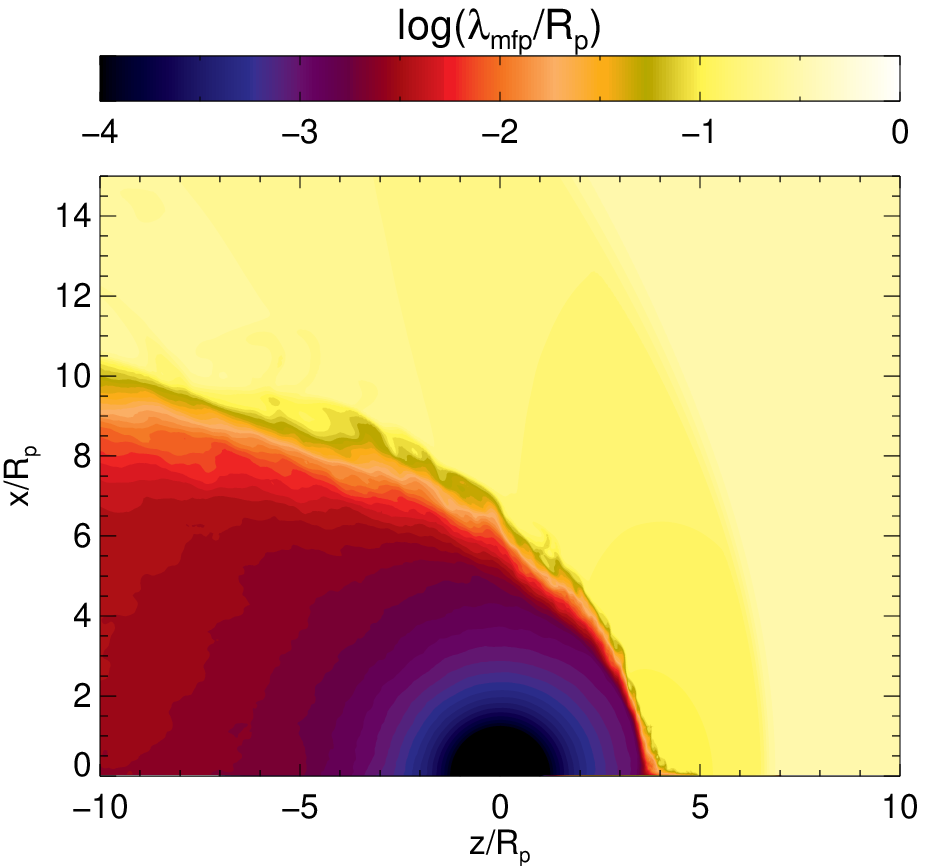}
\caption{The charge-exchange mean free path for a cold hydrogen atom for the case of $\lambda=5$, in units of the planetary radius.   }
\label{Fig:CEMFP}
\end{figure}

To gauge the applicability of the fluid approximation, we calculate the charge-exchange
mean free paths throughout the computational domain.    For a charge exchange rate 
$\nu_{\rm Hp} \approx \beta\left(n_{\rm p,hot}+n_{\rm p, cold}\right)$,  the charge
exchange mean free path for a hydrogen atom is approximately $\lambda_{\rm mfp} =
c_{\rm s}/\nu_{\rm Hp}$ where $c_s$ is the characteristic thermal
sound speed.   For the cold population, $c_{\rm s} \approx 13\,{\rm
  km/s}$, with the sound speed within the hot population an order of
magnitude larger.  

Figure \ref{Fig:CEMFP} shows the charge-exchange mean free path for a
cold hydrogen atom.  Within the region dominated by planetary gas, the
mean free paths are significantly smaller than the
planetary radius, and only become comparable to
the planetary radius in the stellar wind gas.   We can thus be
satisfied with the fluid approximation within the planetary gas.

While the charge exchange rate is taken as constant here for simplicity,
it does exhibit a
temperature dependence depending on the temperatures of the neutral
($T_{\rm n}$) and ionized ($T_{\rm p}$) populations considered.  The
mean free path for a hydrogen atom scales as $\lambda_{\rm mfp}
\propto \left(T_{n}/\left(T_n+T_p\right)\right)^{1/2}$
\citep{2004iono.book.....S}.  For neutral and ionized species with the same temperature, hot or cold, 
the result is shown in Figure \ref{Fig:CEMFP}. If the neutral is hot and the ion is cold,
the mean free path is $\sqrt{2}$ times larger than in Figure \ref{Fig:CEMFP}, and the 
qualitative result is still that the mean free paths are smaller than a planetary radius.
If the neutral is cold and the ion hot, the mean free path is 10 times smaller than shown
in the plot, even more in the hydrodynamic limit.

\red{While the simulations conducted are non-magnetic, both the planetary and stellar wind flows are likely magnetized.  Within the ionized gas of the stellar wind, the gyro-radius for protons moving at the sound speed $r = m_p c_s c/eB \sim 10^2\,{\rm cm} \left(c_s/10\, {\rm km\, s^{-1}}\right)\left(1\, {\rm G}/B\right)$ is significantly smaller than the relevant length scales, thus the motion of the ionized gas can be acceptably treated as a fluid (see \citealt{2015A&A...578A...6M} for a more complete discussion).  }.

\section{ Lyman $\alpha$ transmission spectrum }
\label{sec:lymanalpha}

There are many factors that influence the absorption and scattering of stellar Lyman $\alpha$ photons by atomic hydrogen: the size and shape of the ``cloud" surrounding the planet, column density through the cloud, and the thermal and  bulk velocity of the atoms. The goal of this section is to translate the density and velocity profiles for cold and hot hydrogen into the observable change in the stellar Lyman $\alpha$ spectrum. Position-dependent absorption over the stellar disk will be discussed, as well as disk-integrated absorption versus time during the transit. Comparison of models with different $\lambda$, as well as tilting, ``by hand",  the direction of the stellar wind with respect to the star (to mimic the effect of planetary orbital motion), enables a discussion of how binding energy and geometrical effects are imprinted on the stellar spectrum.

\begin{figure}
\epsscale{1.2}
\plotone{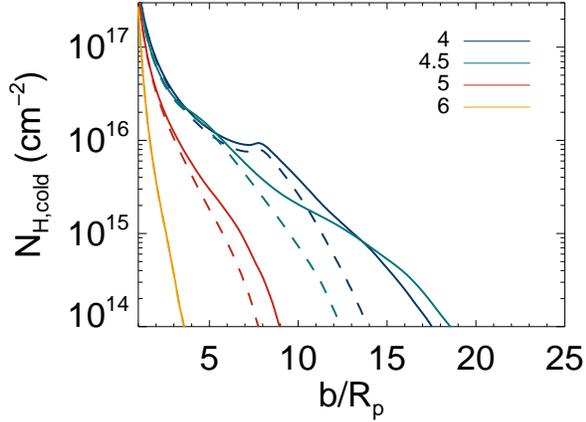}
\caption{Column of cold hydrogen versus impact parameter. Each line is for a different value of $\lambda$. The solid lines integrate over the $z$ coordinate over the entire grid, while the dashed lines only integrate over $z \ge -20R_{\rm p}$.}
\label{Fig:ColdHColumnVsImpact}
\end{figure}

\begin{figure}
\epsscale{1.2}
\plotone{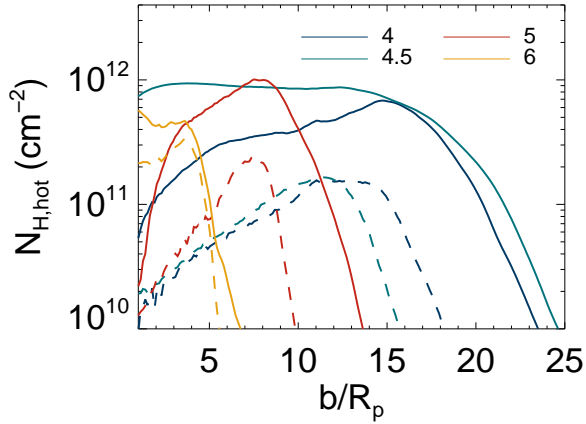}
\caption{Column of hot hydrogen versus impact parameter. Each line is for a different value of $\lambda$.
The solid lines integrate over the $z$ coordinate over the entire grid, while the dashed lines only integrate over  $z \ge -20R_{\rm p}$.}
\label{Fig:HotHColumnVsImpact}
\end{figure}

The column density
\begin{equation}
N_{\rm H}(x,y)  = \int_{-\infty}^\infty dz\ n_{\rm H}(x,y,z),
\end{equation}
 along with the temperature-dependent cross section, is a key quantity in the optical depth.  The relative importance of the cold versus hot hydrogen can be
understood through Figures \ref{Fig:ColdHColumnVsImpact} and \ref{Fig:HotHColumnVsImpact}, 
which show the column of cold and hot hydrogen species as a function of impact parameter.  To assess the distribution along the line of sight, the solid lines show an integration over the entire range of $z$ at a given impact parameter $b$, while the dashed lines are limited to the region with  $z \ge -20\ R_{\rm p}$.  As
  expected, the cold hydrogen is concentrated at small impact 
parameters near the planet. The hot hydrogen exists as a thin layer at the substellar interface (see the dashed lines in Figure \ref{Fig:HotHColumnVsImpact}) with the column increasing with impact parameter.   Downstream,  the hot hydrogen forms a sheath around the planetary gas, ultimately becoming mixed into the tail, resulting in increased columns (see the solid lines in Figure \ref{Fig:HotHColumnVsImpact}).  The cold hydrogen column is seen to extend to larger impact parameter for smaller $\lambda$, due to the larger scale height and stronger outflow from the planet. The deviation of the solid and dashed lines is modest for the cold hydrogen, and again reflects the size of the ``bubble" blown into the stellar wind. The column of hot hydrogen in Figure \ref{Fig:HotHColumnVsImpact} is more complicated. For large $\lambda$, the column is small at small impact parameters, reflecting the fact that the column near the substellar point is smaller than that in the sheath extending toward the tail. Also, the fact that the dashed and solid lines differ indicates significant hot hydrogen column in the tail downstream, as compared to the column near the substellar point. 

Position-dependent absorption at coordinates $(x,y)$ on the stellar disk is calculated as the fractional decrease in flux,
\begin{equation}
\frac{\delta F}{F}(x,y)=\frac{\int d\lambda e^{-\tau_{\rm ISM}\left(\lambda\right)}\left(1-e^{-\tau_{\rm H}\left(x,y; \lambda\right)}\right)I_\lambda^\star}{\int d\lambda e^{-\tau_{\rm ISM}\left(\lambda\right)}I_\lambda^\star}\,\, ,
\label{Eqn:dFF}
\end{equation}
integrated over all wavelengths. 

Here the profile for the stellar Lyman-$\alpha$ line, $I_\lambda^\star$, is proportional to
\begin{equation}
I_\lambda^\star \propto \exp\left(-\frac{1}{2}\left(\frac{\Delta v}{64\,{\rm km\, s^{-1}}}\right)^2 \right)\frac{c}{\lambda^2}
\end{equation}
as in \citep{2010A&A...514A..72L}. The velocity $\Delta v = c \Delta \lambda/\lambda$ is the distance from line center in velocity units.
The optical depth of the interstellar medium $\tau_{\rm ISM}$ is calculated using the cross section $\sigma_{\rm Ly\alpha}$ assuming a mean temperature of $T_{\rm ISM} = 8000\,{\rm K}$ and a column of neutral hydrogen of $N_{\rm H, ISM} = 10^{18.3}\,{\rm cm^{-2}}$, the standard value adopted for HD 209458 \citep{2005ApJS..159..118W},

\begin{equation}
\tau_{\rm ISM}\left(\lambda \right) = \sigma_{\rm Ly\alpha}\left(\lambda; T=8000\,{\rm K}\right) N_{\rm H,ISM}\,\, .
\end{equation}

The optical depth $\tau_{\rm H}$ along a ray $(x,y)$ is given by the line integral
\begin{eqnarray}
\tau_{\rm H}\left(\lambda \right) & = & \int dz\ \left[ n_{\rm H,hot}\sigma_{\rm Ly\alpha}\left(\lambda; T=10^6\,{\rm K},v_{\rm los}\right)\right.\nonumber \\
 & & + \left. n_{\rm H,cold}\sigma_{\rm Ly\alpha}\left(\lambda; T=10^4\,{\rm K},v_{\rm los}\right)\right]\,\, .
\end{eqnarray}
The Lyman-$\alpha$ cross-section $\sigma_{\rm Ly\alpha}$ is given by
\begin{equation}
\sigma_{\rm Ly\alpha}\left(\lambda;T,v_{\rm los}\right) = \frac{\pi e^2}{m_ec} \frac{f_{12}}{\sqrt{\pi}\Delta \nu_{\rm D}}H\left(a,u\right)
\end{equation}
\noindent where $f_{\rm 12} = 0.42$ is the oscillator strength for the $1s\rightarrow 2p$ transition, $\Delta \nu_{\rm D}=\left(\nu_0/c\right)\sqrt{2k_{\rm B}T/m_{\rm H}}$ is the Doppler width,  and $H\left(a,u\right)$ is the Voigt function.   The damping parameter is $a = A_{\rm 21}/(4\pi \Delta\nu_D)$ and the distance from line center in Doppler widths is 
\begin{equation}
u = \frac{\nu-\nu_0}{\Delta \nu_D}+\frac{v_{\rm los}}{v_{\rm th}}\,\, .
\end{equation}
The Einstein A value for Lyman $\alpha$ is $A_{21}=6.3\times 10^8\ {\rm s^{-1}}$.
Limb darkening or brightening are ignored.  

\begin{figure*}
\epsscale{1.0}
\plotone{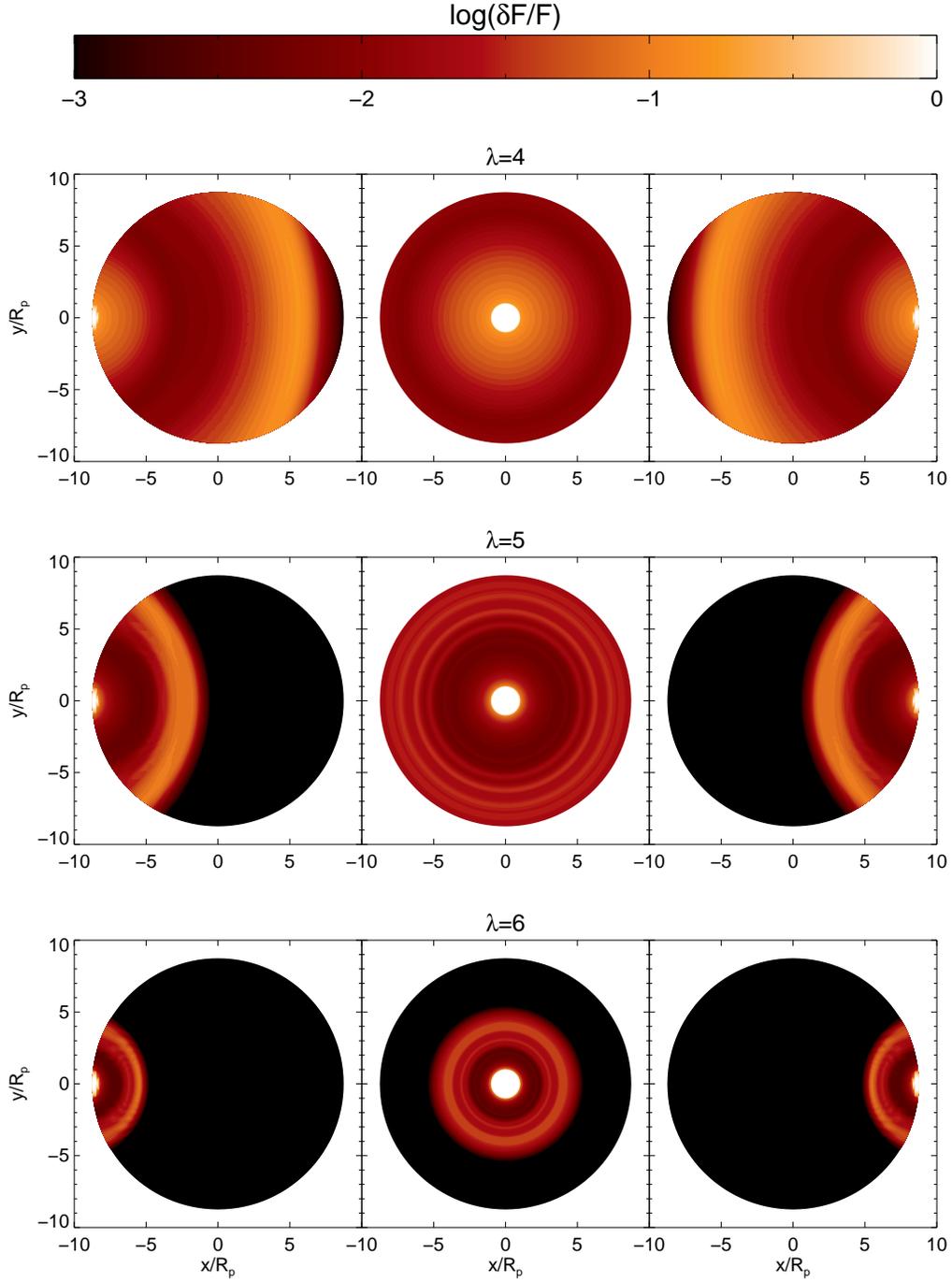}
\caption{The absorption (as defined in eqn. \ref{Eqn:dFF}) with a fixed tail orientation, $\theta_{\rm tail}=0\degree$, directly away from the star. Each row corresponds to a value of $\lambda$, labeled above the panel. Within each row, the three plots are ingress, mid-transit, and egress, from left to right.
}
\label{Fig:TransitThree00}
\end{figure*}

The Lyman-$\alpha$ absorption is computed at three orbital phases, ingress, mid-transit and egress, corresponding to an orbital phase
  angle of $-7.5\degree$, $0\degree$, and $+7.5\degree$, respectively, and for three orientations of the tail relative to the star-planet axis, $\theta_{\rm tail} = 0\degree,\, 20\degree,\, 40\degree$. The value $\theta_{\rm tail} = 0\degree$ corresponds to the tail pointing from the planet directly away from the star, while $\theta_{\rm tail} = \, 20\degree$ and 
  $40\degree$ denote a tail lagging behind the planet's motion. Here ingress and egress are for the optical continuum; the transit durations are longer observed at Lyman $\alpha$ wavelengths compared to \red{the} optical continuum.  \red{For typical parameters for close-in planets, the tail angle is likely to be significantly offset from $\theta_{\rm tail} = 0\degree$; however, we include the $\theta_{\rm tail} = 0\degree$ case for comparison.}

Figure \ref{Fig:TransitThree00} shows the position-dependent absorption
 across the face of the star for the case of the tail pointing
radially away from the star ($\theta_{\rm tail} = 0\degree$, the 
aligned case).
The opaque central disk with $\delta F/F = 1$ is the optical continuum transit radius, covering a fractional
area $\simeq 1\%$ of the star (radius $R_\star=1.2R_\odot$, see Table \ref{table:209458b}). The dark outer regions are transparent at Lyman $\alpha$ wavelengths, with absorption $\delta F/F \leq 10^{-3}$. In this $\theta_{\rm tail} = 0\degree$ case, the plots for ingress and egress are symmetric about the center of the star. More of the star is covered by high absorption sightlines for the smaller $\lambda=4$ case, due to the larger mass loss rate and columns.  

\begin{figure*}
\epsscale{1.0}
\plotone{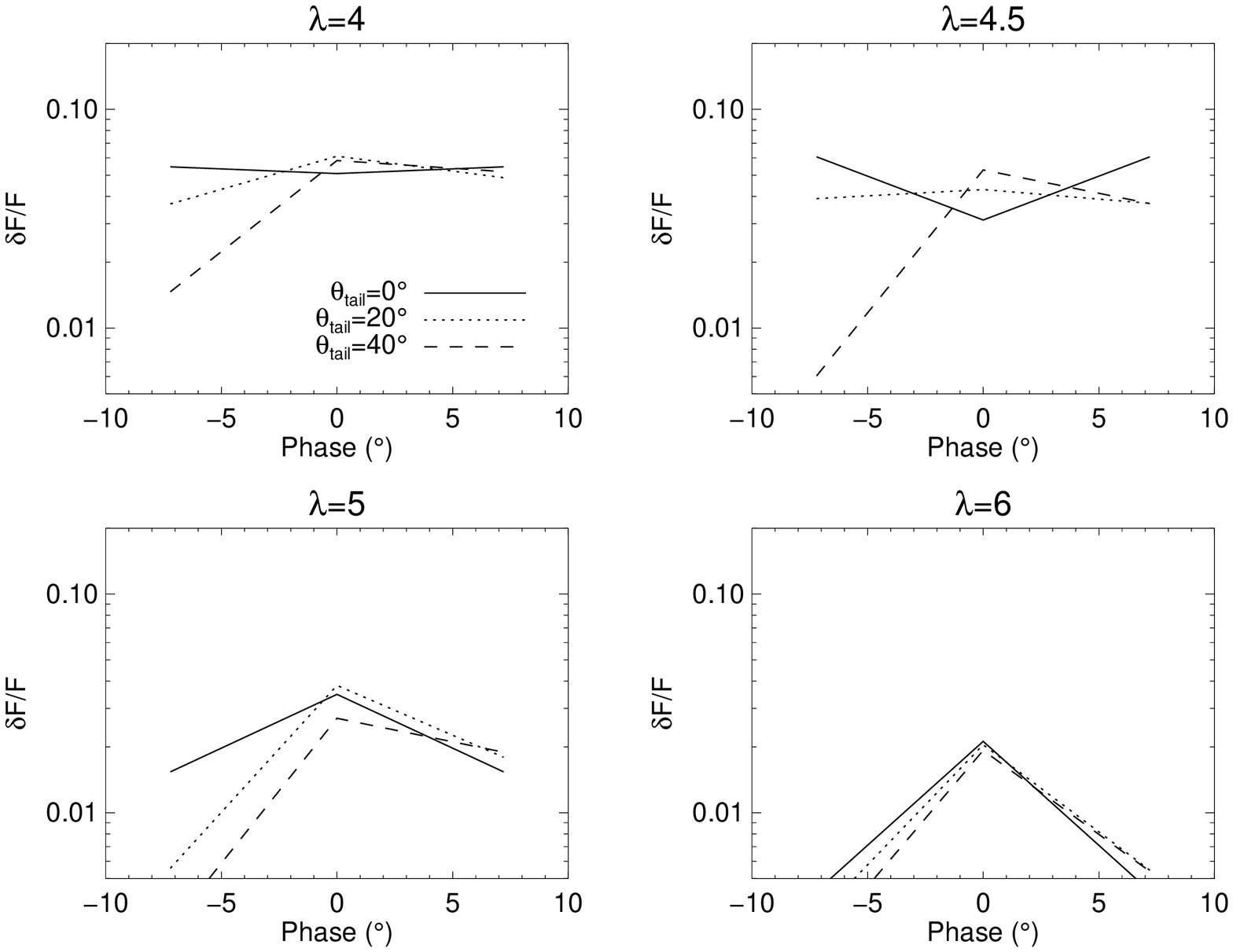}
\caption{The total absorption as a function of orbital phase -- ingress, mid-transit, egress - for four different values of $\lambda$ and three orientations of the tail: $\theta_{\rm tail} = 0\degree,\,\,20\degree,\,\,40\degree$.}
\label{Fig:RelativePhase}
\end{figure*}

The absorption has a ring profile around the planet which is most
apparent in the case of $\lambda=6$ in Figure
\ref{Fig:TransitThree00}, as the entire tail fits inside the stellar disk for this case (see Figures \ref{Fig:HotHContour} and \ref{Fig:ColdHContour}). The ringed appearance can be understood as absorption from cold atoms decreasing away from the planet, while absorption from hot atoms increases away from the planet, over the impact parameter range of interest. The cold hydrogen absorption is more important at small $\lambda$, as the scale height is larger. The hot population is much more constant with impact parameter, by comparison. 
While the sheath of hot
hydrogen exists for all $\lambda$,  it becomes larger than the disk of the star as $\lambda$ decreases, making it unobservable in absorption
at mid-transit. In cases where the majority of the sheath isn't seen mid-transit, it
can still be observed at ingress and egress.  For small $\lambda$,
this leads to an increased absorption at ingress and egress compared
to the mid transit (see Figure \ref{Fig:RelativePhase}). For large
$\lambda$, the reverse is seen, with absorption largest at mid-transit.

\begin{figure*}
\epsscale{1.0}
\plotone{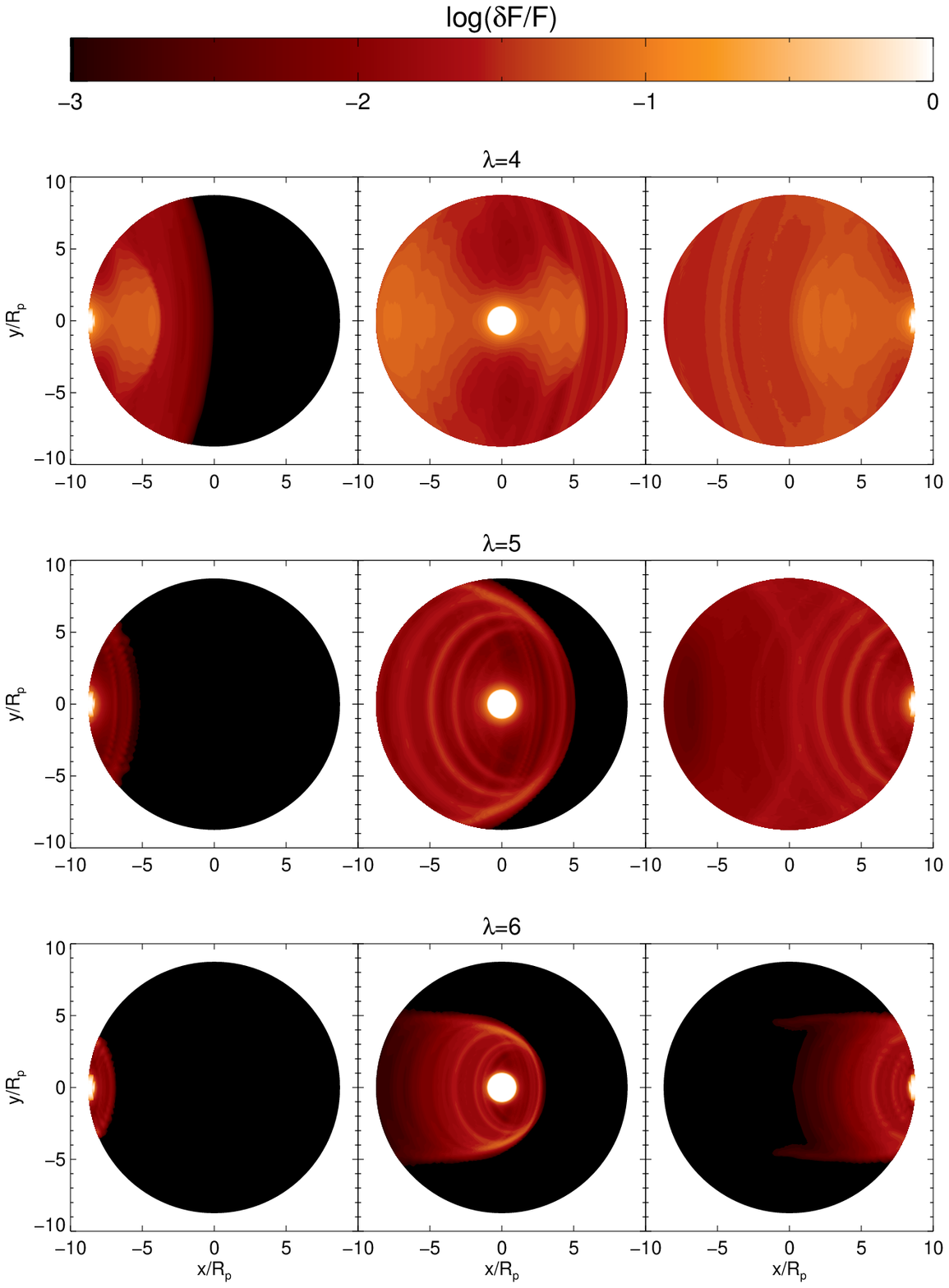}
\caption{The absorption (as defined in eqn. \ref{Eqn:dFF}) with a fixed tail orientation $\theta_{\rm tail}=40\degree$ for three values of $\lambda$.}
\label{Fig:TransitThree40}
\end{figure*}

Figure \ref{Fig:TransitThree40} shows the effect of a stellar wind misaligned with the star-planet line by $\theta_{\rm tail} = 40\degree$. The misalignment angle is much larger than the offset orbital phase angle at (optical continuum) ingress and egress, and so the tail is always strongly pointing to the left in the plots. The same axisymmetric simulations are used for the density and velocity profiles to make this plot, but the rays from the star are assumed to travel through this distribution at the given angle. The effect of a misaligned trailing tail reduces the absorption at ingress while increasing it at egress as the
trailing gas continues to occult a significant portion of the stellar
disk (see Figure \ref{Fig:TransitThree40}).

The disk integrated absorption, integrated over all frequencies, is computed  from Equation \ref{Eqn:dFF} as
\begin{eqnarray}
\frac{\delta F}{F} & =& 
\frac{ \int d\lambda\ e^{-\tau_{\rm ISM}(\lambda)} 
         \int dxdy\ \left( 1 - e^{-\tau_{\rm H}(x,y;\lambda) } \right) I_\lambda^\star }
         {\pi R_\star^2 \int d\lambda\ e^{-\tau_{\rm ISM}(\lambda)} I_\lambda^\star }
         \label{eq:dFoverF}
\end{eqnarray}
Figure \ref{Fig:RelativePhase} shows the integrated $\delta F/F$ versus orbital phase, evaluated at the optical ingress, mid-transit, and optical egress for four different values of $\lambda$. For the $\lambda=4$ case, the hot hydrogen distribution is so large that in the aligned case the transit depth is nearly constant over the entire optical transit. For the case of a misaligned tail, the depth is smaller at ingress than egress, as expected. The other extreme case of $\lambda=6$ shows the transit depth decreases strongly away from mid transit as the tail is much smaller than the stellar disk. The transition cases $\lambda=4.5$ and $5$ are more complicated during the transition between the two regimes.

\begin{figure}
\epsscale{1.0}
\plotone{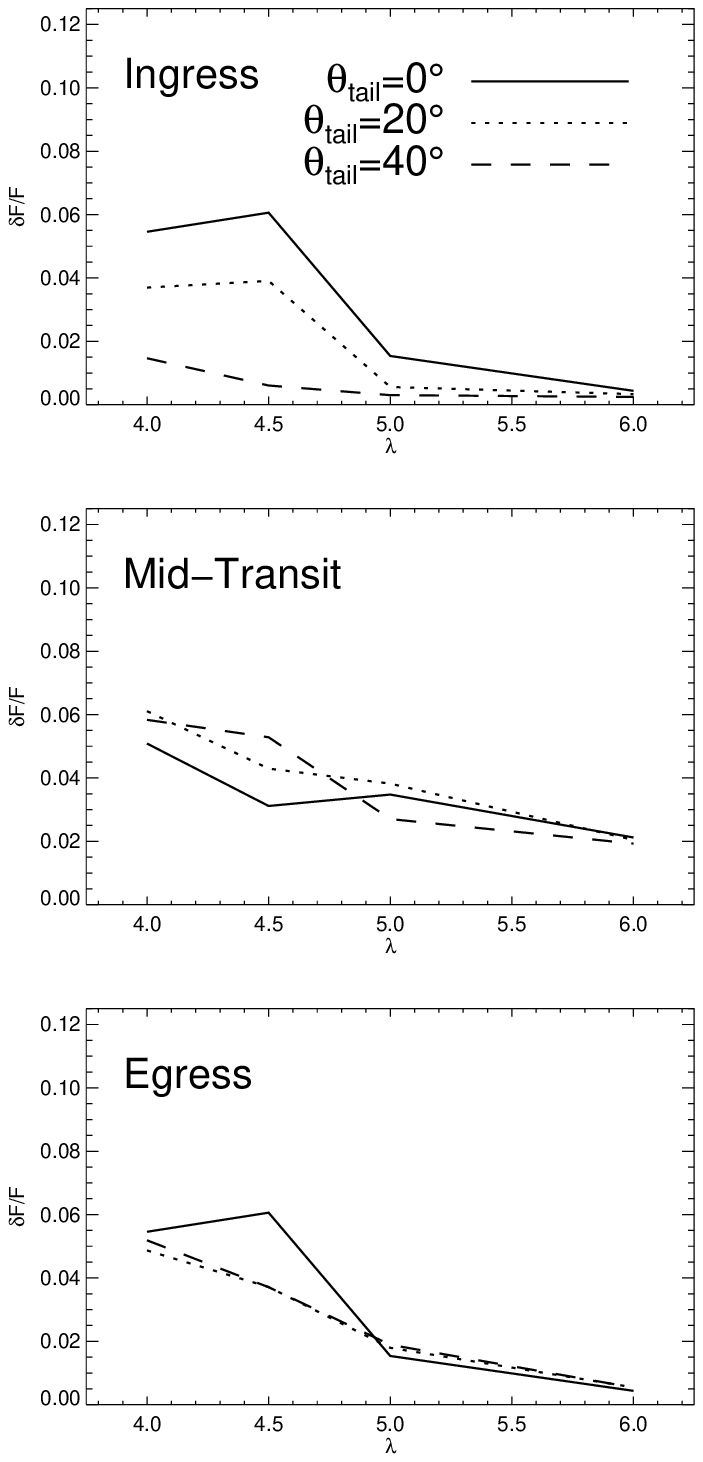}
\caption{The total absorption as a function of $\lambda$ at ingress (top), mid-transit (middle), and at egress (bottom) for three values of $\theta_{\rm tail}$.}
\label{Fig:dFFvsLambda}
\end{figure}

Figure \ref{Fig:dFFvsLambda} shows the integrated transit depth for the three different tail orientations, now as a function of $\lambda$. Aside from $\lambda=4.5$, there is a general trend of less absorption with increasing $\lambda$, by a factor of a few to 10 over the given range of $\lambda$. The transition case $\lambda=4.5$ case has more hot hydrogen at impact parameters covering the stellar disk, leading to increased absorption compared to $\lambda=4$ or $\lambda=5$ (see Figure \ref{Fig:HotHColumnVsImpact}).

\section{Summary and Discussion}
\label{sec:discussion}

The Zeus code was used to perform hydrodynamical simulations of escape from hot Jupiter atmospheres and the subsequent interaction with the stellar wind. Parameters appropriate for HD 209458b and the host star were used. The problem is outlined in \S \ref{sec:outline}, and the numerical method was discussed in \S \ref{sec:numericalmethod}. The use of a polytropic equation of state with $\gamma=1.01$ allowed the $T=10^4\ {\rm K}$ planetary gas and the $T =10^6\ {\rm K}$ stellar wind gas to remain separately isothermal, except in the mixing layer between the two gases where intermediate temperatures were reached. While the stellar wind was held fixed, the surface temperature of the planet, parametrized through $\lambda$, was varied in order to change the strength of the outflow from the planet. 

Results for the total density and velocity are shown in Figures \ref{Fig:InstVsAvg}, \ref{Fig:DensContour}, \ref{Fig:VelocityPlots} and \ref{Fig:SpeedContour}. For small $\lambda=4$, a ``colliding winds" structure is found, as in \citet{2013MNRAS.428.2565T}. The planetary gas accelerates outward to supersonic speeds and ``blows a bubble" in the stellar wind, comparable in size to the stellar disk. As the supersonic planetary gas decelerates before meeting the stellar wind, a high density post-shock shell is created. A modest increase to $\lambda=6$ gives a remarkably different structure than the $\lambda=4$ case. The falloff of the density in the hydrostatic portion of the atmosphere is much steeper due to the smaller scale height. Also, the stellar wind is able to penetrate into the planet's atmosphere deeper than where the sonic point would have been (Figure \ref{Fig:VelocityPlots}), and as a result there is no transonic outflow from the planet. This is referred to as the ``viscous" regime. Compared to the $\lambda=4$ case, the $\lambda=6$ case shows a much narrower tail, with large densities extending to only a couple planetary radii (Figure \ref{Fig:DensContour}). The velocity contours shown in Figure \ref{Fig:SpeedContour} only show significant planetary gas velocities for the $\lambda=4$ case, while for $\lambda \geq 4.5$ the velocities are always $\la 20-30\rm km\ s^{-1}$. Also apparent in this figure is the large shear layer separating the planetary and stellar wind gas. Instabilities in this layer will eventually lead to mixing of the planetary and stellar wind gases sufficiently far downstream, as seen in the instantaneous values in Figure \ref{Fig:InstVsAvg}. The mixing layer is close behind the planet for $\lambda=6$, while for $\lambda=4$ the mixing occurs further downstream.

Figure \ref{Fig:MassLossResStudy} shows mass loss as a function of $\lambda$. For small $\lambda \la 5$, mass loss is entirely set by the surface boundary condition on the planet, and is independent of the stellar wind parameters, agreeing well with the analytic formula for mass loss in an isothermal wind. Here $\lambda$ is a proxy for the amount of stellar irradiation heating the atmosphere, and hence small $\lambda$ corresponds to a planet closer to the star. For $\lambda \ga 5$, an abrupt decrease in the mass loss rate is seen when the stellar wind penetrates inside the sonic point of the planetary outflow (Figure \ref{Fig:VelocityPlots}). The ``turbulent" mass loss formula from \citet{1982MNRAS.198.1007N}, in which turbulent viscosity generated in the shear layer between the two fluids is able to entrain and accelerate planetary gas, is shown to give mass loss rates accurate to a factor of 2 over a range of $\lambda=5.5-8.5$. In this regime, the surface temperature of the planet (implicitly set by stellar irradiation) determines the scale height of the gas, and the atmosphere is truncated when the stellar wind pressure is balanced by planetary gas pressure. Contrary to the colliding winds regime, mass loss is sensitive to stellar wind parameters in the viscous regime.

Given an understanding of the basic hydrodynamic flow, the behavior of the hot and cold populations of neutral \red{hydrogen} and protons can be discussed. From \S \ref{sec:numericalmethod}, each of the four species is advected with the hydrodynamic flow just discussed. Charge-exchange between hydrogen atoms and protons were included, allowing the creating of hot hydrogen, and the inverse reactions. As discussed by \citet{2013MNRAS.428.2565T}, the timescale for these reactions is so short they rapidly come into rate equilibrium in the mixing layer. Also included in the present study are optically thin photo-ionization and electron-impact collisional ionization by stellar wind electrons. These two reactions act as sinks for the atoms, and are crucial to limit the extent of the neutral cloud surrounding the planet. For the fiducial values used here, collisional ionization is a factor of 10 larger than photoionization, although it occurs only in the mixing layer.

Figure \ref{Fig:InstVsAvg} shows the instantaneous versus time-averaged density contours for cold and hot hydrogen for the $\lambda=4.5$ case. The hot hydrogen distribution is far more variable than that of cold hydrogen due to eddies in the mixing layer. Figures \ref{Fig:HotHContour} and \ref{Fig:ColdHContour} show time-averaged density profiles for the hot and cold hydrogen, respectively, for a range of $\lambda=4-6$. While the cold hydrogen is always centered on the planet, and decreases much more strongly away from the planet as $\lambda$ is increased, the hot hydrogen distribution is more complicated. In the $\lambda=4$ colliding winds case, the hot hydrogen forms a sheath around the planetary wind. While the density is larger on the leading edge of the flow (to the right in the plots), the leading edge is very thin. The hot hydrogen sheath surrounding the tail has smaller densities, but the path length through the sheath is far larger than the leading edge, and hence the tail may be a promising site for Lyman $\alpha$ absorption. For the $\lambda=6$ viscous case, mixing occurs immediately behind the planet and hot hydrogen fills the tail region just behind the planet, unlike the $\lambda=4$ case. The hot hydrogen tail is also much narrower, only blocking a fraction of the stellar disk.

Figure \ref{Fig:CEMFP} is a self-consistency check on the fluid approximation used in this paper for hydrogen atoms (see e.g. \citealt{2009ApJ...693...23M} for a discussion of mean free paths). This figure shows the mean free path for a cold hydrogen atom to undergo a charge exchange reaction with either a cold or hot proton. The mean free path for the $\lambda=5$ case is found to be much smaller than other lengthscales inside the planetary gas, verifying that the fluid approximation is valid there. In the stellar wind gas, the mean free path approaches $\sim (0.1-1) \times R_{\rm p}$ and is comparable to the flow lengthscales. A more accurate prescription would allow ballistic motion of hydrogen atoms over mean free paths $\sim (0.1-1) \times R_{\rm p}$ in the stellar wind between interactions. 

Translation of the spatial distribution of cold and hot hydrogen into transit depths is investigated through plots of the column depth in Figures \ref{Fig:ColdHColumnVsImpact} and \ref{Fig:HotHColumnVsImpact} and the frequency integrated absorption $\delta F(x,y)/F$ along each sightline to the star in Figures \ref{Fig:TransitThree00} and \ref{Fig:TransitThree40}. As expected, the cold hydrogen column density and absorption decrease away from the planet. This is due to cold hydrogen density decreasing as the gas expands geometrically outward, \red{charge exchange reactions, and ionization}. This outward decrease appears as a central disk in the aligned tail case in Figure \ref{Fig:TransitThree00}. By contrast, the hot hydrogen column in Figure \ref{Fig:HotHColumnVsImpact} shows a more complicated structure. 
Focusing on the solid lines for integration over the entire grid, the column is not strongly peaked at small impact parameter. Either it is constant at small impact parameter before falling off, or it rises to a peak at impact parameters $5-15\ R_{\rm p}$. In Figure \ref{Fig:TransitThree00}, this shows up as an annulus of high absorption, falling off to larger or smaller impact parameter. When considering both cold and hot hydrogen, a minimum (dark ring in Figure \ref{Fig:TransitThree00}) is then expected, as absorption by cold hydrogen decreases outward to a minimum, then increases due to absorption by hot hydrogen, and then decreases again at large impact parameter. 

We emphasize it is the sheath of hot hydrogen around the tail, and not the thin layer of hot hydrogen at the leading edge that gives rise to the ring of absorption seen in Figure \ref{Fig:TransitThree00}. It was to investigate the contribution of the tail that motivated us to augment the simulations of \citet{2013MNRAS.428.2565T} by using spherical geometry and ionization reactions. 

Lastly, the frequency and disk-integrated transit depth $\delta F/F$ defined in Equation \ref{eq:dFoverF} is shown as a function of orbital phase in Figure \ref{Fig:RelativePhase} and as a function of $\lambda$ for Figure \ref{Fig:dFFvsLambda}. Three different viewing sightlines through the tail are computed to simulate the effect of the planet's orbital motion, which would cause the tail to be swept backward in the orbit, as opposed to radially away from the star. Due to the cold (absorption decreasing outward from the planet) plus hot (absorption in an annulus) ring structure of the absorption, the aligned case shows little change in Lyman $\alpha$ transit depth during the entire optical transit for the $\lambda=4$ and $4.5$ cases. For the $\lambda=5$ and $6$ cases, the absorption falls off more steeply from the planet, and the absorption is then peaked about mid-transit. When the tail is misaligned, pointing backward in the orbit, there is less absorption at ingress and more absorption at egress, as expected. The Lyman $\alpha$ transit duration would be significantly longer than the optical transit for the $\lambda=4$ and $4.5$ cases. As expected, the transit depth decreases with increasing $\lambda$, due to smaller planetary scale heights and weaker mass outflow. Only the transition case $\lambda=4.5$ deviates from this, likely due to the filled in structure at small impact parameter in  Figure \ref{Fig:HotHColumnVsImpact}.

While we have investigated the role of the parameter $\lambda$ on the mass loss rate and the dynamics of the star-planet interaction, a number of other possible parameters have been held fixed.   Increasing (decreasing) the density at the base of the atmosphere, $\rho_{\rm p,b}$, will increase (decrease) the mass loss rate and $\lambda_c$ (see Eq. \ref{Eqn:lambdac}).  Since the transit depth depends on not only the density of the neutral gas but also the size of the sheath of hot hydrogen relative to the size of the host star, changes in $\rho_{\rm p,b}$ can result in non-trivial changes in the transit depth.   Similarly, the stellar wind parameters $n_\star$, $e_\star$ and $v_\star$ have remained fixed.   Varying these parameters will also result in commensurate variations in $\lambda_c$ and can alter the production of hot hydrogen in the interaction layer. 

By adopting a near-isothermal equation of state with spherical launching of the planetary wind,  we do not capture the effects of asymmetric irradiation (e.g., \citealt{2009ApJ...694..205S,2015arXiv150606759T}) which can result in nightside accretion and could affect the dynamics of the tail.   Our simulations, due to their axisymmetry, also only allow for the tail to point directly away from the star and do not include stellar gravity or tidal forces.    These can result in the deformation of the tail,  exposing it more directly to the stellar wind.   \red{Stellar and tidal forces can also provide additional acceleration to the wind, increasing the possibility that a supersonic wind develops.}

We have also neglected the effect of the planetary magnetic field which could potentially limit the outflow \citep{2011ApJ...728..152T,2014ApJ...788..161T,2014MNRAS.444.3761O} and could limit mixing of the hot and cold populations \citep{2013MNRAS.428.2565T}.   \red{The magnetic field could alter the structure of shock, and the possibility of observing magnetic bow shocks has been a subject of considerable investigation (e.g., \citealt{Vidotto2010,Llama2011,llama2013})}

We found the mean free paths to be small in the planetary gas but to be become comparable to characteristic planetary scales within the stellar wind.  This signals the breakdown of the fluid treatment in the stellar wind.  The larger mean free paths in the stellar wind could allow for the interaction layer with a spatial extent greater than is found in our simulations.  
The short mean free paths in the planetary gas have implications for Direct Simulation Monte Carlo (DSMC) models which
neglect planetary ions (e.g., \citealt{2014Sci...346..981K}).  By 
ignoring the planetary ions, which are the dominant impactor, neutral hydrogen atoms attain potentially unphysical velocities by radiation pressure and charge exchange.   \citet{2014Sci...346..981K} also include a region of exclusion through which stellar ions cannot pass\footnote{\citet{2014Sci...346..981K} launch the wind from an inner boundary using parameters characteristic of one dimensional hydrodynamic models.  Even at the inner boundary, the gas would be significantly ionized (see \citealt{2013Icar..226.1678K}) which would result in ions coupled to the magnetic field providing a source of drag to inhibit the launching of a wind.}. This is intended to be interpreted as deflection by the planetary magnetosphere; however, this region would be populated with planetary ions which will exert a drag force on any accelerated neutrals.   Furthermore, escaping ionized gas from the planet is capable of creating a similar obstacle without the requirement of a magnetic field as can be seen from our simulations.    In the limit where the obstacle is purely hydrodynamic in origin (e.g., the case of our simulations), the transit signal is expected to be maximized, as for a similarly sized obstacle, magnetic pressure replaces thermal pressure, requiring less gas capable of absorbing Lyman-$\alpha$ from the star.  

Future investigations done in the fluid limit should include acceleration of the gas by Lyman-$\alpha$ radiation to determine the degree to which the the stellar irradiation is able to accelerate neutral gas which is well coupled to the ionized planetary gas. 

% \appendix
 
\acknowledgements

The authors thank Chenliang Huang and Shane Davis for valuable conversations.	 This work was supported by NASA Origins of Solar Systems Grants NNX14AE16G and NNX10AH29G.  Zhi-Yun Li is supported in part by NSF AST-1313083.
\bibliographystyle{apj}
\bibliography{winds}

\label{lastpage}
\end{document}